\documentclass[12pt,a4paper]{article}
\usepackage{graphicx}
\usepackage{amssymb}
\usepackage{amsmath}
\usepackage{amsthm}
\def\fund{{  \> {\vcenter  {\vbox  
              {\hrule height.6pt
               \hbox {\vrule width.6pt  height5pt  
                      \kern5pt 
                      \vrule width.6pt  height5pt }
               \hrule height.6pt}
                         }
                   }
           \>\> }}

\def\fundbar{{  \> \overline{ {\vcenter  {\vbox  
              {\hrule height.6pt
               \hbox {\vrule width.6pt  height5pt  
                      \kern5pt 
                      \vrule width.6pt  height5pt }
               \hrule height.6pt}
                         }
                   } }
           \>\> }}

\def\symm{{  \> {\vcenter  {\vbox  
              {\hrule height.6pt
               \hbox {\vrule width.6pt  height5pt  
                      \kern5pt 
                      \vrule width.6pt  height5pt 
                      \kern5pt
                      \vrule width.6pt height5pt}
               \hrule height.6pt}
                         }
              }
           \>\> }}

\def\symmbar{{  \> \overline{ {\vcenter  {\vbox  
              {\hrule height.6pt
               \hbox {\vrule width.6pt  height5pt  
                      \kern5pt 
                      \vrule width.6pt  height5pt 
                      \kern5pt
                      \vrule width.6pt height5pt}
               \hrule height.6pt}
                         }
              }
           } \>\> }}

\def\asymm{{ \> {\vcenter  {\vbox  
              {\hrule height.6pt
               \hbox {\vrule width.6pt  height5pt  
                      \kern5pt 
                      \vrule width.6pt  height5pt }
               \hrule height.6pt
               \hbox {\vrule width.6pt  height5pt  
                      \kern5pt 
                      \vrule width.6pt  height5pt }
               \hrule height.6pt}
                         }
              }
           \>\> }}

\def\asymmbar{{ \> \overline{ {\vcenter  {\vbox  
              {\hrule height.6pt
               \hbox {\vrule width.6pt  height5pt  
                      \kern5pt 
                      \vrule width.6pt  height5pt }
               \hrule height.6pt
               \hbox {\vrule width.6pt  height5pt  
                      \kern5pt 
                      \vrule width.6pt  height5pt }
               \hrule height.6pt}
                         }
              }
           } \>\> }}

\newtheorem{theorem}{Theorem} 

\newcommand{\rap}[2]
{\setbox1=\hbox{#1}%
\setbox2=\hbox to\wd1{\hss #2\hss}%
\mbox{\rlap{\box1}\box2}}

\newcommand{\sla}[1]{\rap{$#1$}{/}}
\newcommand{\oplane}[4]{\textstyle {\genfrac{[}{]}{0pt}{}{{\scriptstyle #4}\  {\scriptstyle #3}}{{\scriptstyle #1} \ {\scriptstyle #2}}}}

\newcommand{\oplaneone}[1]{\oplane{ {#1}_1}{ {#1}_2}{ {#1}_3}{ {#1}_4}}

\newcommand{\oplanewide}[4]{{[\!\!\tiny  \begin{array}{c@{} c c} #4 &\, & #3 \\ #1 & \, &#2 \end{array}\!\!]}}
\newcommand{\oplanetwo}[2]{\oplanewide{ {#1}_1 {#2}_1}{ {#1}_2 {#2}_2}{ {#1}_3 {#2}_3}{ {#1}_4 {#2}_4}}

\def\tr{\mathop{\rm tr}\nolimits}
\def\mod{\mathop{\rm mod}\nolimits}
\def\notni{\mathop{\sla\ni}\nolimits}

\newcommand{\ol}{\overline}
\newcommand{\wt}{\widetilde}

\newcommand{\B}{{{\cal C}^*}}

\newcommand{\bC}{\ensuremath{\mathbb{C}}}
\newcommand{\bZ}{\ensuremath{\mathbb{Z}}}
\newcommand{\bT}{\ensuremath{\mathbb{T}}}

\newcommand{\balpha}{\mbox{\boldmath $\alpha$}}
\newcommand{\bbeta}{\mbox{\boldmath $\beta$}}

\theoremstyle{plain}
\newtheorem{Rule}{Rule} % \rule already defined

\def\since{%
\setlength{\unitlength}{1pt}%
\thinlines %
\begin{picture}(6,6)%
\put(0 ,4){.}
\put(2.5 , 0){.}
\put(5,4){.}
\end{picture}%
~
}%

\begin{document}
%%%%%%%%%%%%%%%%%%%%%%%%%%%%%%%%%%%%%%%%%%%%%%%%%%%%%%%%%%%%%%%%%%%%%%%%

\begin{titlepage}
\title{\hfill\parbox{4cm}
       {\normalsize UT-07-37\\January 2008}\\
       \vspace{1.5cm}
Anomalies and O-plane charges\\
in orientifolded brane tilings
       \vspace{1.5cm}}
\author{
Yosuke Imamura,\thanks{E-mail: \tt imamura@hep-th.phys.s.u-tokyo.ac.jp}\quad
Keisuke Kimura\thanks{E-mail: \tt kimura@hep-th.phys.s.u-tokyo.ac.jp}\\
and\\
Masahito Yamazaki\thanks{E-mail: \tt yamazaki@hep-th.phys.s.u-tokyo.ac.jp}%
\\[20pt]
{\it Department of Physics, University of Tokyo,}\\ {\it Tokyo 113-0033, Japan}
}
\date{}

\maketitle
\thispagestyle{empty}

\vspace{0cm}

\begin{abstract}
\normalsize
We investigate orientifold of brane tilings.
We clarify how the cancellations of
gauge anomaly and Witten's anomaly are
guaranteed by the conservation of the D5-brane charge.
We also discuss the relation between
brane tilings and the dual Calabi-Yau cones
realized as the moduli spaces of gauge theories.
Two types of flavor D5-branes in
brane tilings and
corresponding superpotentials of fundamental quark fields
are proposed, and
it is shown that the massless loci of these quarks in the
moduli space correctly reproduce the worldvolume of
flavor D7-branes in the Calabi-Yau cone dual to the fivebrane system.
\end{abstract}

\end{titlepage}

\section{Introduction}
\begin{figure}[htb]
\centerline{\scalebox{0.5}{\includegraphics{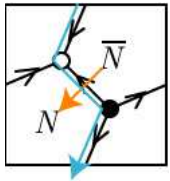}}}
\caption{The bipartite graph of $\bC^3$ is shown.
This diagram has two vertices of opposite colors,
three edges from black to white, and one hexagonal face.
Among three arrows representing three adjoint fields and three
zig-zag paths representing the boundary of semi-infinite cylinders
of NS5-branes,
one for each is shown.}
\label{orientation.eps}
\end{figure}

Brane tilings\cite{Hanany:2005ve,Franco:2005rj,Franco:2005sm} are two-dimensional diagrams drawn on tori which are used to represent
the structure of a large class of quiver gauge theories.
They can be used to describe arbitrary ${\cal N}=1$ superconformal
field theories realized on D3-branes in toric Calabi-Yau cones.
They are dual graphs
of corresponding quiver diagrams, and
vertices and edges in a graph represent $SU(N)$ factors in
the gauge group and chiral multiplets belonging to
bi-fundamental representations.
The graphs are bipartite.
Namely, vertices are colored black and white, and
any pair of vertices connected by an edge have different
colors.
The orientation of edges are determined according to
the colors of vertices at the endpoints.
We take it from black to white.
Because $SU(N)$ factors correspond to faces,
it is natural to represent a bi-fundamental field $\Phi^a{}_b$
as an arrow connecting two adjacent faces for two $SU(N)$ factors coupling to
$\Phi^a{}_b$.
The orientation of arrows, which specifies one of
$(N,\ol N)$ or $(\ol N,N)$, is also determined with the
colors of vertices.
We take the convention that if the orientation of
an edge is South to North
the orientation of the arrow intersecting with the edge
is East to West.
The head and tail of an arrow correspond to the upper color index for
the fundamental representation and the lower color index for the anti-fundamental
representation, respectively (Figure \ref{orientation.eps}).

The important feature of brane tilings is
that they are directly related to the topological
structure of fivebrane systems
realizing the gauge theories, as is clarified in \cite{Feng:2005gw}.
We can regard the torus on which a bipartite graph is drawn
as a stack of D5-branes, and edges in the bipartite graph
as NS5-branes intersecting with the D5-branes.
We can reconstruct the worldvolume of the NS5-brane by
attaching semi-infinite cylinders to the torus along
zig-zag paths in the diagram.
A zig-zag path is a path made of edges in a bipartite graph
defined so that when we go along the path we choose the leftmost
edge at white vertices and the
rightmost edge at black vertices.
We can regard
a bipartite graph
as the superposition of
zig-zag paths.
Because every edge is included in
two zig-zag paths, each edge can be regarded as
a part of the intersection of D5-branes and NS5-branes.

We label four dimensions in which the gauge theory lives
by $0123$, and the two cycles in the torus by $5$ and $7$.
The orientation of branes in the system is shown in
Table \ref{branes.tbl}.
\begin{table}[htb]
\caption{The structure of fivebrane systems represented by brane tilings.}
\label{branes.tbl}
\begin{center}
\begin{tabular}{c|cccc|cccc|cc}
\hline
\hline
& $0$ &$1$& $2$& $3$& $4$& $5$& $6$& $7$& $8$& $9$ \\
\hline
D5 & $\circ$ & $\circ$ & $\circ$ & $\circ$ & & $\circ$ && $\circ$ \\
NS5 & $\circ$ & $\circ$ & $\circ$ & $\circ$ & \multicolumn{4}{c|}{$\Sigma$} \\
\hline
\end{tabular}
\end{center}
\end{table}
$\Sigma$ in the table is a two-dimensional non-compact surface
in $4567$ space.
All branes are in the subspace $x^8=x^9=0$, and the system
possesses the rotational symmetry on $89$-plane.
This is an R-symmetry in the gauge theory,
which is not necessarily the R-symmetry in the superconformal group
unless it is appropriately mixed with the gauge symmetries
on the fivebrane system.
In the weak string coupling limit the tension of NS5-branes becomes infinitely larger
than that of D5-branes, and the worldvolume of the NS5-brane
becomes a smooth holomorphic surface in the $4567$ space.
It is given by
\begin{equation}
P(e^{x^4+ix^5},e^{x^6+ix^7})=0,
\end{equation}
where $P(u,v)$ is the Newton polynomial
associated with the toric diagram of the toric Calabi-Yau cone.
The NS5-brane worldvolume has branches going to infinity on the $46$-plane.
Each branch of NS5-brane is topologically semi-infinite cylinder, which is
attached on
the D5-branes along a zig-zag path.
The asymptotic structure of NS5-brane projected on the $46$-plane
gives the web-diagram of the Calabi-Yau, while the brane tiling
can be regarded as the projection of NS5-branes into the $57$-plane.
This brane system is related with the system
of D3-branes in the toric Calabi-Yau cone
by T-duality along the $57$ directions.
It is possible to obtain information about gauge theories
such as anomalies, marginal deformations, etc.
by studying corresponding fivebrane systems\cite{Imamura:2006ub,Imamura:2006ie,Imamura:2007dc}.

We can generalize the brane systems by introducing extra ingredients.
The introduction of fractional D3-branes in the Calabi-Yau set-up
can be realized in the fivebrane system by
assigning integers representing D5-brane charges to external lines
in the web-diagram\cite{Butti:2006hc,Imamura:2006ub}.
This changes each asymptotic part of the NS5-brane
into D5-NS5 bound state.
The D5-brane charge conservation requires the numbers of D5-branes
on faces change depending on the numbers assigned to external lines,
and this gives different ranks of $SU(N)$ factors in the gauge group.
It can be shown that the gauge anomaly cancels
when the D5-brane charge conserves\cite{Imamura:2006ub}.

We can also introduce flavor branes, which give fields
belonging to the fundamental and the anti-fundamental representations.
In \cite{Franco:2006es} non-compact D7-branes wrapped on divisors in
toric Calabi-Yau cones are investigated, and the matter contents
which these flavor branes give rise to are proposed.
Such flavor D7-branes are dual to D5-branes spreading along
012346 directions in the fivebrane systems.
We discuss this type of flavor branes in \S\ref{d7.sec}
and extend the results in \cite{Franco:2006es}.

Recently, orientifolds of brane tilings are investigated in \cite{Franco:2007ii}.
There are several possibilities of orientifold planes
which preserve ${\cal N}=1$ supersymmetry.
In \cite{Franco:2007ii}, O5-planes, fixed points in bipartite graphs,
and O7-planes, fixed lines in graphs, are investigated,
and simple prescription to obtain
field contents are given.
The orientations of these orientifold planes are given in Table \ref{oplanes.tbl}.
\begin{table}[htb]
\caption{The structure of fivebrane systems with orientifolds. As shown in this table, both O5-planes and O7-planes preserve $\mathcal{N}=1$ supersymmetry. In this paper we only consider the case of O5-planes.}
\label{oplanes.tbl}
\begin{center}
\begin{tabular}{c|cccc|cccc|cc}
\hline
\hline
& $0$ &$1$& $2$& $3$& $4$& $5$& $6$& $7$& $8$& $9$ \\
\hline
D5 & $\circ$ & $\circ$ & $\circ$ & $\circ$ & & $\circ$ && $\circ$ \\
NS5 & $\circ$ & $\circ$ & $\circ$ & $\circ$ & \multicolumn{4}{c|}{$\Sigma$} \\
O5 & $\circ$ & $\circ$ & $\circ$ & $\circ$ & $\circ$ && $\circ$ & \\
O7 & $\circ$ & $\circ$ & $\circ$ & $\circ$ & $\circ$ &&& $\circ$ & $\circ$ & $\circ$ \\
\hline
\end{tabular}
\end{center}
\end{table}
They also propose a set of rules which
uniquely determines the $\bZ_2$ orientifold parity
of mesonic operators when the positions of fixed points
in the bipartite graphs and their ``charges'' are specified.
In this paper, we concentrate on the orientifold with O5-planes,
and derive the these rules from the viewpoint of fivebrane systems.
We leave the O7-plane case for future work.

If we construct the gauge theory for an orientifolded fivebrane system
by the naive orientifold projection,
the daughter theory, the theory obtained by the projection
from the parent theory, in general possesses
gauge anomalies.
This can be cured by introducing an appropriate number of
fundamental or anti-fundamental representation fields,
which we call quarks.
A purpose of this paper is to explain how these extra quark fields
arise from the viewpoint of fivebrane system.
By analogy to the relation between the gauge anomaly cancellation
and the D5-brane charge conservation in the un-orientifolded case,
it is natural to expect the
emergence of the fundamental representation in the orientifolded
brane tilings is also guaranteed by the D5-brane charge conservation.
We will show that this is actually the case.

For this to happen it is important that
O5-planes carry the D5-brane charge, and its signature
changes when it intersects with NS5-branes\cite{Evans:1997hk,Hanany:2000fq,Hyakutake:2000mr,Bergman:2001rp}.
As is shown in \S\ref{anom.sec} this property of O5-planes explains how the quark fields arise.
At the same time, this fact raises a problem.
When we use the rules proposed in \cite{Franco:2007ii}
we have to specify ``charges'' of the four fixed points.
These charges, however, cannot be identified with the
RR-charges of orientifold planes because an O5-plane
carries both positive and negative RR-charges if
it intersects with NS5-branes.
We need to distinguish the RR-charges of O5-planes
and ``charges'' used in the rules.
In this paper,
in order to distinguish them from RR-charges of O5-planes,
we refer to the ``charges'' in the rules
as ``transposition parities'' or, simply, ``T-parities'',
because they are directly related to the
symmetry of the corresponding field under the transposition\footnote{
Unfortunately, this name is the same as the T-parity in the little Higgs models\cite{Cheng}. We hope no confusion will arise.}.
To clarify the relation between the T-parity
and the RR-charge of O5-planes
is another purpose of this paper.

This paper is organized as follows.
In the next section, we briefly review the rules
proposed in \cite{Franco:2007ii}.
In \S\ref{anom.sec} we discuss anomaly cancellations,
and we there give an explanation for the
emergence of (anti-)fundamental fields which cancels
the gauge anomaly from the viewpoint of fivebrane systems.
In \S\ref{cy.sec} we discuss the relation
between $\bZ_2$ parity of mesonic operators
and the RR-charges of O5-planes, which may depend on the position
on the O5-planes.
This gives rules to determine $\bZ_2$ parity from the RR charges
of O5-planes.
By comparing these rules and those in \cite{Franco:2007ii}
we relate the RR-charge
and the T-parity.
We investigate the relation between flavor branes and quark fields
in \S\ref{d7.sec}, and propose superpotentials
which correctly reproduce the worldvolumes of flavor D7-branes
in the Calabi-Yau cones
as the loci in which quarks become massless.
\S\ref{discussions.sec} is devoted for discussions.

%%%%%%%%%%%%%%%%%%%%%%%%%%%%%%%%%%%%%%%%%%%%%%%%%%%%%%%%%%%%%%%%%
\section{T-parity and mesonic operators}\label{tparity.sec}
In this section we briefly review the rules to determine
the theories realized on orientifolded brane tilings
proposed in \cite{Franco:2007ii}.
We here only discuss orientifold with O5-planes,
which are represented as four fixed points on the torus.

In the parent theory, each face represents an $SU(N)$ factor in the
gauge group.
If a face is identified with another face by the orientifolding,
we should identify the corresponding two $SU(N)$.
Let $SU(N)$ and $SU(N)'$ be a pair of two factors identified.
The reversal of orientation of open strings implies that
these two factors should be identified via the charge conjugation.
Namely, upper and lower indices of $SU(N)$ correspond
to lower and upper indices of $SU(N)'$, respectively.
The bi-fundamental field $\Phi^a{}_{b'}$ with one upper $SU(N)$ index $a$
and one lower $SU(N)'$ index $b'$,
which exists
if two faces identified are adjacent with each other,
is regarded as field $\Phi^{ab}$ with two upper $SU(N)$ indices
or $\Phi_{a'b'}$ with two lower $SU(N)'$ indices.
The symmetry for two indices of these fields is
determined according to the following rule
proposed in \cite{Franco:2007ii}.
\begin{Rule}[Edge rule]
If a fixed point with positive/negative
T-parity is on an edge,
the field associated with the edge
belongs to the symmetric/antisymmetric
representation. \label{edge.rule}
\end{Rule}
If a face is identified with itself by the orientifolding,
the corresponding gauge group $SU(N)$ becomes
$Sp$ or $SO$ group according to the rule\cite{Franco:2007ii}:
\begin{Rule}[Face rule]
If a fixed point with positive/negative T-parity is inside a face,
$SO$/$Sp$ gauge group lives on the face. \label{face.rule}
\end{Rule}
Because the orientifold flip exchanges white and black vertices,
fixed points cannot be at vertices.

The two rules above are rules for fields
which are mapped to themselves
by the orientifold flip.
The orientifold transformations of
other elementary fields,
which are mapped to other fields,
depend on how we define the relative phases of fields.
In \cite{Franco:2007ii},
instead of giving such transformations for elementary fields,
they give rules for gauge invariant mesonic operators,
which are defined as the trace of product of bi-fundamental fields.
On the brane tiling, such mesonic operators are described as
closed paths made of arrows corresponding to
the constituent bi-fundamental fields.
The $\bZ_2$ parity of a mesonic operator is determined by
combining the following rules:
\begin{Rule}[Product rule]
The $\bZ_2$ parity of a mesonic operator corresponding to
$\bZ_2$ symmetric path passing through two fixed points
      is the product of the T-parities of the fixed points.
\label{product.rule}
\end{Rule}

\begin{figure}[htbp]
\centerline{\scalebox{0.5}{\includegraphics{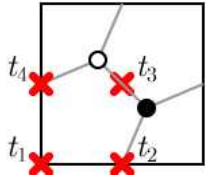}}}
\caption{In the O5-plane case, we have four fixed points on $\bT^2$.
The notation $T=\protect\oplane{t_1}{t_2}{t_3}{t_4}$
is meant to represent T-parities of four orientifold planes,
as shown in this figure.}
\label{tttt.eps}
\end{figure}

\begin{Rule}[Superpotential rule]
  The $\bZ_2$ parity of a mesonic operator appearing in the superpotential
is negative. \label{superpotential.rule}
\end{Rule}
For later convenience we introduce the following expression for product rule:
\begin{equation}
P[{\cal O}]=\int_CT,
\label{rule1}
\end{equation}
where $T=\oplane{t_1}{t_2}{t_3}{t_4}$ are the set of four
T-parities of four O5-planes on $\bT^2$, whose positions are shown in Figure \ref{tttt.eps},
and $\int_C$ represents the product of two
parities assigned to fixed points passed through by a $\bZ_2$ symmetric path $C$.
In \S\ref{cy.sec}, we clarify the relation between T-parity and RR-charge
by using $\bZ_2$ parity obtained by these rules.

As a corollary of the rules above, we can show that
the following sign rule
imposed on the T-parities of the orientifold planes\cite{Franco:2007ii}.
\begin{Rule}[Sign rule]
The product of all the T-parities is equal to $(-1)^{N_W/2}$
where $N_W$ is the number of the terms in the superpotential,
which is equal to the number of vertices in the bipartite graph.
\label{sign.rule}
\end{Rule}

%%%%%%%%%%%%%%%%%%%%%%%%%%%%%%%%%%%%%%%%%%%%%%%%%%%%%%%%%%%%%%%
\section{Anomaly cancellation}\label{anom.sec}
With fivebrane systems described by brane tilings,
we can realize quiver gauge theories with
different ranks depending on faces
by changing the numbers of D5-branes depending on faces.
The D5-brane charge conservation requires that when
the numbers of two adjacent faces are not the same,
the difference must be canceled by the
inflow of the charge from the NS5-brane corresponding to edges.
In the un-orientifolded case, it can be shown\cite{Imamura:2006ub}
that the gauge anomaly cancels
if the brane system is consistent with the D5-brane charge conservation law.

The purpose of this section is to show that this is the case
for orientifolded brane tilings.
The gauge group of an orientifolded theory consists of
$SU(N)$, $SO(N)$, and $Sp(N/2)$ factors.
(When we consider gauge group $Sp(N/2)$, we always assume that $N$
is an even integer.)
If a face does not have
fixed point inside it or on its boundary
the anomaly cancellation are guaranteed
in the same way as the un-orientifolded case by
the D5-brane charge conservation.
In the following we discuss anomaly cancellation for
a face with an O5-plane inside it or on its boundary.

\subsection{O5-planes inside faces}\label{face.sec}
If a face has a fixed point inside it and
is identified with itself by the orientifold projection,
the gauge group realized on the face is $SO$ or $Sp$.
These groups do not have the ordinary gauge anomaly.
We should, however, take care of the Witten's anomaly\cite{Witten:1982fp}
for $Sp$ gauge groups.
The cancellation of Witten's anomaly requires
the number of fundamental representations for each $Sp$ group
must be even.
We can easily show that this condition automatically holds if
the D5-charge conservation law is satisfied in the brane system.

\begin{figure}[htb]
\centerline{\scalebox{0.5}{\includegraphics{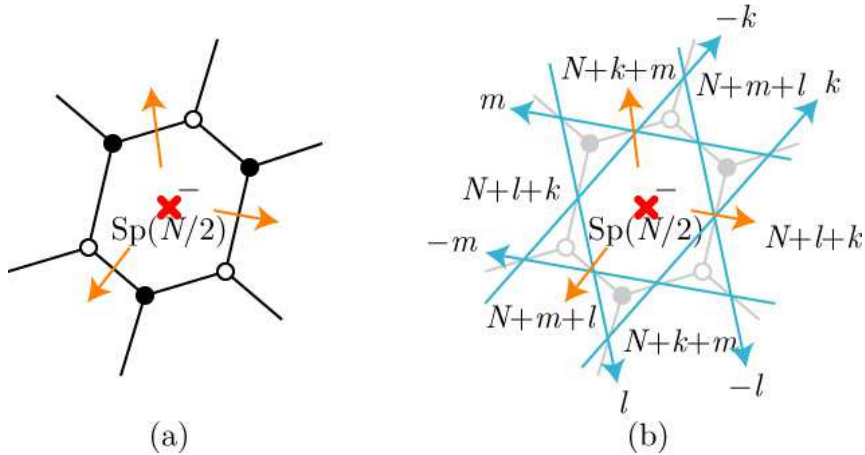}}}
\caption{A face with fixed point inside it.
The independent bi-fundamental fields are shown as outgoing arrows.}
\label{spface.eps}
\end{figure}

Because black vertices are mapped to white vertices
by the orientifold flip,
a face with O5-plane inside it has $4n+2$ edges.
For concreteness let us consider an example of a hexagonal face
shown in Figure \ref{spface.eps} (a).
Generalization to other cases is trivial.
If the T-parity of the fixed point is negative
$Sp(N/2)$ gauge group lives on
the hexagonal face at the center.
The face is enclosed by six edges.
Only three of them are independent and the other three are mirror images.
When we consider anomaly, only bi-fundamental fields for the independent three
edges should be taken into account.
We can, for example, take three outgoing arrows as independent bi-fundamental fields.

When we discuss fivebrane charge conservation,
it is convenient to use diagrams
in which zig-zag paths are represented as smooth cycles
going on the right/left side of black/white vertices
in the zig-zag path.
(Figure \ref{spface.eps} (b))
We call such diagrams {\em fivebrane diagrams}.
(These are the same as what are referred to as ``rhombus loop diagrams'' in \cite{Hanany:2005ss})

\begin{figure}[htb]
\centerline{\scalebox{0.5}{\includegraphics{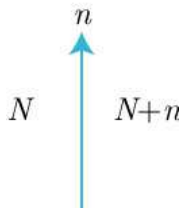}}}
\caption{The D5-brane charges assigned to two adjacent faces
and the edge between them are shown.}
\label{inflow.eps}
\end{figure}

When the numbers of D5-branes on faces are different,
the difference of the D5-branes charge must be
supplied by the attached NS5-branes,
which are represented as cycles in the diagram.
In order to assign charges to faces
in a consistent way with the D5-brane charge conservation,
we first assign D5-charges to cycles,
and the charges of faces are determined so that
if we go across
a cycle from one face to another,
the D5-charges of the faces change by the charge assigned to the
cycle.
In our convention, if up-going cycle carries charge $n$
and $N$ is assigned to the face on the left side,
the number assigned to the right is $N+n$.
(Figure \ref{inflow.eps})
Two cycles which are mirror to each other must have the
same D5-charges with opposite sign because
the orientifold flip reverses the orientation of the cycles.
The charges of six adjacent faces are shown in Figure \ref{spface.eps}.
The numbers of $Sp(N/2)$ fundamental representations for
three independent arrows are
\begin{equation}
N_1=N+k+m,\quad
N_2=N+m+l,\quad
N_3=N+l+k.
\end{equation}
Because $N$ is an even integer,
the total number $N_1+N_2+N_3=3N+2k+2m+2l$
is also an even integer,
and the Witten's anomaly does not arise.

%%%%%%%%%%%%%%%%%%%%%%%%%%%%%%%%%%%%%%%%%%%
\subsection{O5-planes on edges}\label{edge.sec}
When an orientifold plane is on an edge of
a bipartite graph, two $SU(N)$ gauge groups
on both sides of the edge are identified,
and the bi-fundamental field coupling to these two
gauge groups becomes symmetric or antisymmetric
representation of the $SU(N)$ gauge group.
The anomaly coefficient $d_R$ for these tensor representations
are given by
\begin{equation}
d_\asymm=(N-4)d_\fund,\quad
d_\symm=(N+4)d_\fund.
\end{equation}
These are different from the contribution $Nd_\fund$ of
the bi-fundamental field in the parent theory,
and we need extra ingredient in order to cancel
the gauge anomaly.
The simplest way to cancel this anomaly is
to introduce four fundamental or anti-fundamental
chiral multiplets as is pointed out in \cite{Franco:2007ii}.

How these new matter fields arise in the brane system?
A natural way to introduce these (anti-)fundamental fields
is to introduce four flavor branes.
If we introduce four D5-branes coinciding with the O5-plane,
we may obtain four fundamental fields.
(This cannot be directly shown by quantizing open strings
due to the complicated structure of the brane system.)
However, this answer is not satisfactory.
On the gauge theory side, we must introduce
four (anti-)fundamental representations so that
the anomaly cancels.
Otherwise the theory would be inconsistent.
On the other hand, at first sight, it seems possible
to introduce an arbitrary number of flavor branes.
Note that we do not have to require the cancellation
of RR-charge carried by the O5-plane.
Because some of transverse directions of the O5-plane
is non-compact, the RR-flux induced by O5-plane and flavor branes
can escape to infinity.
The RR-flux may cause the breaking of conformal symmetry,
but it does not cause any inconsistency at all.

Moreover, the RR-charge of O5-plane (which are defined
as integral of flux over $RP^3$, not over $S^3$) is $\pm1$,
and the number of D5-branes including mirror images
required to cancel the
O5-charge is $2$.
Even if we introduced flavor D5-branes
which cancel the O5-charge, we do not obtain
the desired number of fundamental representations.

The key to solve this puzzle is
the fact that in our brane system
O5-planes and NS5-branes co-exist,
and when an O5-plane intersects with NS5-branes
it changes its RR-charge\cite{Evans:1997hk,Hanany:2000fq,Hyakutake:2000mr,Bergman:2001rp}.
In the brane system we consider here
an O5-plane is a two dimensional plane in 4567 space.
Its worldvolume is spread along non-compact 46 directions.
The NS5-branes are also two dimensional surfaces,
and if corresponding cycle, zig-zag path, on the tiling goes through
the fixed point, it shares one direction with
the O5-plane.

\begin{figure}[htb]
\centerline{\scalebox{0.5}{\includegraphics{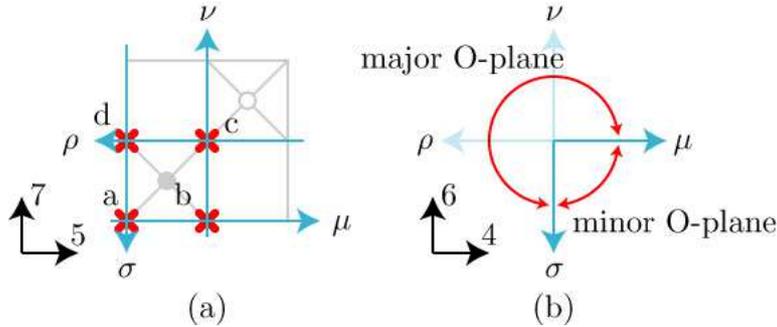}}}
\caption{The example of the conifold. Here we show $\bT^2$ (57) directions.
$\mu$, $\nu$, $\rho$, $\sigma$ are cycles of NS5-branes and $a,b,c,d$ are intersection points of O5-planes with D5.}
\label{conifold.eps}
\end{figure}

Let us consider the conifold case as an example.
The fivebrane diagram is given in Figure \ref{conifold.eps} (a).
$a$, $b$, $c$, and $d$ are O5-planes
and the cycles $\mu$, $\nu$, $\rho$, and $\sigma$ are NS5-branes.
Each of them spreads along the following directions in 4567 space:
\begin{equation}
a,b,c,d:46,\quad
\mu,\rho:45,\quad
\nu,\sigma:67.
\end{equation}
We see that, for example, the O5-plane $a$ and the NS5-brane $\mu$ share one direction $x^4$,
and they intersect along a line.
The O5-plane $a$ also intersects with NS5-brane $\sigma$
along a line.
As a result, the O5-plane $a$ is divided into two parts
by the two NS5-branes $\mu$ and $\sigma$.
On the 46-plane, these two parts are represented as
one quadrant and the rest
(Figure \ref{conifold.eps} (b)).
\begin{figure}[htb]
\centerline{\scalebox{0.5}{\includegraphics{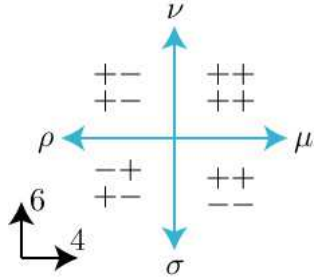}}}
\caption{RR-charges of four O5-planes.
RR-charge assignment
changes when crossing NS5-brane cycles ($\mu,\nu,\rho,\sigma$)
and thus depends on quadrants on 46-plane.}
\label{conifold2.eps}
\end{figure}
In general an O5-plane at an intersection of two cycles in
a fivebrane diagram is divided into two parts by
two legs in the web-diagram.
We call these two parts minor and major O5-planes
according to their central angles.
Because the RR-charge of the O5-plane
changes when it intersects with NS5-branes,
the minor and major O5-planes for the same fixed point
have opposite RR-charges to each other.
This is the case for other three orientifold planes, $b$, $c$, and $d$
in the conifold example,
and the RR-charge assignments to the four orientifold planes
depend on quadrants on 46-plane.
One example is shown in Figure \ref{conifold2.eps}.

\begin{figure}[htb]
\centerline{\scalebox{0.5}{\includegraphics{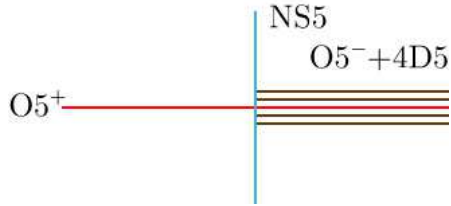}}}
\caption{When O5-brane intersects with NS5-brane, its RR-charge changes its sign.Moreover, in order to conserve RR-charge, inclusion of four flavor D5-branes (including their mirrors) on top of O5-plane are required.}
\label{ns5o5.eps}
\end{figure}

Now let us take account of the RR-charge conservation.
The simplest way to satisfy the conservation law is
to introduce four (including mirror images)
D5-branes on top of O5$^-$-plane compensating the
change of O5-plane's RR-charge at the intersection of
O5 and NS5 (Figure \ref{ns5o5.eps}).
With the assumption of the gauge anomaly cancellation,
we expect one of the following two combinations arise
at the O5-plane.
\begin{equation}
\mbox{positive T-parity} : \asymm + 4\fund~,\quad
\mbox{negative T-parity} : \symm + 4\fundbar~.
\end{equation}

More generally
we can introduce more flavor branes
spreading over whole 46-plane.
It is also possible to transfer the excess of RR charge
as a flow on the NS5-brane, and put flavor branes off the O5-planes.
These possibilities are discussed in \S\ref{d7.sec}

%%%%%%%%%%%%%%%%%%%%%%%%%%%%%%%%%%%%%%%%%%%%%%%%%%%%%%%%%%%%%%%%%
\section{Relation to Calabi-Yau cones}\label{cy.sec}
\subsection{Orientifold of general toric CY cones}\label{cycone.sec}
The purpose of this subsection is to establish the relation
between the RR charge of O5-planes and the $\bZ_2$ parity
of mesonic operators with the help of
the T-duality between fivebrane systems and Calabi-Yau cones.

\begin{figure}[htb]
\centerline{\scalebox{0.5}{\includegraphics{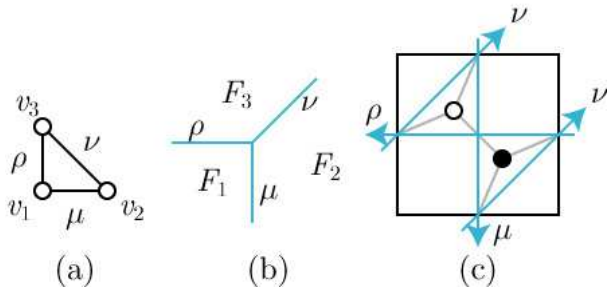}}}
\caption{Shown here are toric diagram (a), web-diagram (b),
and bipartite graph with fivebrane diagram (c) for $\bC^3$.}
\label{c320.eps}
\end{figure}

\paragraph{Toric diagram}
Let us start from a toric Calabi-Yau cone $\cal M$ described by a
toric diagram.
In this paper we only consider three-dimensional Calabi-Yau manifolds.
Let $v_i\in \Gamma=\bZ^3$ be the set of lattice points in
the toric diagram.
By $SL(3,\bZ)$ transformation, we can take the coordinate
system in which the components of $v_i$ are given by
\begin{equation}
v_i=(p_i,q_i,1).\label{vi}
\end{equation}
The toric diagram is
usually represented as
a two-dimensional diagram by using the first two components of these vectors.
An example of $\bC^3$ case is shown in Figure \ref{c320.eps} (a).

We define the dual cone $\B$ as the set of vectors $w\in \mathbb{R}^3$
satisfying
\begin{equation}
v_i\cdot w\geq0\quad
\forall i.
\label{dualcone}
\end{equation}
When we consider resolutions of the toric Calabi-Yau,
K\"ahler parameters come to the right hand side of this inequality.
In this paper we will not discuss such resolutions
and the right hand side is always zero.
In such a case
we do not have to use all $v_i$
to define $\B$ by (\ref{dualcone}),
and we only need $v_\alpha$ corresponding to the corners of
the toric diagram. Here notation $\alpha,\beta \ldots$ is used to denote lattice points in the corners of toric diagram, and $i,j \ldots$ denotes all the lattice points in the toric diagram. We further assume that the label $\alpha$ increases one by one as we go around the perimeter of toric diagram in counterclockwise manner.
The boundary of the dual cone $\partial \B$ consists of flat faces called facets.
Each facet corresponds to each vector $v_{\alpha}$,
and is defined as the set of points satisfying $v_\alpha\cdot w=0$ and $v_{\beta} \cdot w \ge 0 ~~(\forall \beta\ne \alpha)$.
We denote the facet corresponding to $v_\alpha$ by $F_\alpha$.
The structure of the base manifold $\B$ is conveniently expressed as
a planar diagram by projecting the facets onto a two-dimensional
plane by simply neglecting the third coordinate.
It is called a web-diagram.
Figure \ref{c320.eps} (b) is an example of web-diagram for $\bC^3$.
The lines in the web-diagram represent the edges of the
base manifold $\B$.

We can regard the Calabi-Yau manifold as
the $\bT^3$ fibration over the dual cone $\B$ (\ref{dualcone}), although strictly speaking some cycles of $\bT^3$ shrinks on facets as we will explain.
Let $(\phi_1,\phi_2,\phi_3)$ be the coordinates in the toric
fiber.
We choose the period of each coordinate to be $2\pi$.
We can regard $\Gamma$ as the lattice associated with
the toric fiber $\bT^3$.
Namely, we can associate points in $\Gamma$ with cycles in
$\bT^3$.
By this identification, we can regard an arbitrary non-vanishing vector $v\in\Gamma$
as a generator of $U(1)$ isometry
of the $\bT^3$.
We denote the symmetry generated by $v$ by $U(1)[v]$.
Two flavor symmetries which do not rotate the supercharges
are $U(1)[(1,0,0)]$ and $U(1)[(0,1,0)]$,
and R-symmetry is $U(1)[(a_1,a_2,1)]$.
When $a_1$ and $a_2$ are appropriately chosen
this gives the R-symmetry in the superconformal algebra\cite{Martelli:2005tp}.

On a facet $F_\alpha$ the cycle specified by $v_\alpha$ in $\bT^3$ fiber
shrinks and the fiber becomes $\bT^2$.
In order to parameterize the $\bT^2$ fiber on each facet,
the following coordinate change is convenient.
\begin{equation}
(\phi_1,\phi_2,\phi_3)=\theta_1(1,0,0)+\theta_2(0,1,0)+\theta_3(p_i,q_i,1).
\end{equation}
This is equivalent to
\begin{equation}
\theta_1=\phi_1-p_{\alpha}\phi_3,\quad
\theta_2=\phi_2-q_{\alpha}\phi_3,\quad
\theta_3=\phi_3. \label{thetadef}
\end{equation}
On the facet, the $\theta_3$-cycle shrinks and
$(\theta_1,\theta_2)$ is a pair of good coordinates on $\bT^2$.
By taking T-duality along these two angular coordinates
we obtain fivebrane system
described by the brane tiling.
The third angular coordinate $\theta_3$ is identified
with the argument of the complex coordinate $x^8+ix^9$
in the fivebrane system.

\begin{figure}[htb]
\centerline{\scalebox{0.5}{\includegraphics{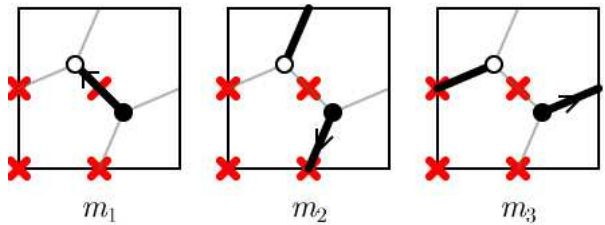}}}
\caption{The three perfect matchings for the bipartite graph of $\bC^3$ are shown.}
\label{c3pm.eps}
\end{figure}

\paragraph{Perfect matchings and zig-zag paths}
A perfect matching is a subset of edges of a bipartite graph
such that for each vertex one (and only one) of the edges ending on the vertex
is included in the subset.
Figure \ref{c3pm.eps} shows the three perfect matchings
for $\bC^3$.
Perfect matchings plays important roles in the connection
between bipartite graphs and Calabi-Yau geometry.
See \cite{kenyon} for review.
Given a perfect matching,
we can define the corresponding unit flow
by regarding each matched edge as black-to-white flow by one.
A unit flow is a flow with source $1$ at each black vertex and
sink $1$ at each white vertex.
Flows corresponding to perfect matchings are special kind of
unit flows.
We abuse the term ``perfect matchings'' in the following to mean
also the corresponding unit flows.
The difference of two unit flows is a conserved flow.
For such conserved flows we can uniquely define the
flux across closed cycles.
For a flow $f$ and a cycle $C$, we denote
the total flux of $f$ across $C$ by $\langle f,C\rangle$.
We define this flux so that if $C$ is an up-going cycle
$\langle f,C\rangle$ is the flux of $f$ passing $C$ from left to right.
The set of two fluxes across $\balpha$ and $\bbeta$-cycles
is called height change,
and we denote it by $h(f)$.
\begin{equation}
h(f)=(\langle f,\balpha\rangle,\langle f,\bbeta\rangle). \label{heightdef}
\end{equation}
(We take the $\balpha$ and $\bbeta$-cycles along $x^5$ and $x^7$,
respectively. Also, in this paper we use bold letters
to denote $\balpha$ and $\bbeta$-cycles in order to distinguish them
from $\alpha, \beta, \ldots$, which are labels of corners of the toric diagram.)

Let us label perfect matchings by indices $\alpha'$, $\beta'$, $\ldots$.
We can define ``relative positions'' between
two perfect matchings $m_{\alpha'}$ and $m_{\beta'}$
by the height change $h(m_{\alpha'}-m_{\beta'})$,
and we can plot matchings as points on a two-dimensional lattice.
An important fact is that this set of points is nothing but the toric diagram \cite{Hanany:2005ve,Franco:2006gc}.
By this fact we can associate each vertex in the toric diagram
with perfect matchings.
In the $\bC^3$ example, the matchings $m_1$, $m_2$, and $m_3$ in Figure \ref{c3pm.eps}
correspond to the facets $F_1$, $F_2$, and $F_3$, respectively of Figure \ref{c320.eps} (b).
(We use the term ``facets'' not only for faces of dual cones
but also for their projection onto the web diagrams.)
In general several perfect matchings may be associated with
one vertex.
For a vector $v_i$ in the toric diagram,
let $m[v_i]$ be one of the associated matchings.
Then the following relation holds:
\begin{equation}
h(m[v_i]-m[v_j])=v_i-v_j.
\label{mviab}
\end{equation}
This equation makes sense because the third component of $v_i-v_j$ always vanishes and
gives a vector in the two-dimensional lattice. Also,
even though we have in general several perfect matchings
associated with vertices $v_i$ and $v_j$,
this equation holds regardless of the choice of perfect matchings
$m[v_i]$ and $m[v_j]$.

There are some arguments that
there are bipartite graphs which do not give
physical quiver gauge theories\cite{Hanany:2005ss}.
To obtain meaningful gauge theories
we need to impose the condition that
there is one and only one perfect matching
associated for each corner in toric diagrams.
In the following we only consider such ``good'' bipartite graphs.
We denote the unique perfect matching for a corner $\alpha$
by $m_\alpha$.
This one-to-one correspondence between corners of the toric diagram
and perfect matchings of the bipartite graph plays important roles in what follows.

For a general vector $v\in \bZ^3$,
we define $m[v]$ in the following way.
First we decompose $v$ into linear combination
of $v_i$ as $v=\sum_i a_i v_i$ with integral
coefficients $a_i$.
Then $m[v]$ is defined by
\begin{equation}
m[v]=\sum_i a_im[v_i]. \label{vdecomp}
\end{equation}
The source at black vertices and sink at white vertices
of the flow $m[v]$ are always the same,
and it is equal to
the sum of the coefficients, $\sum_ia_i$.
This is nothing but the third component of $v$,
and thus independent of how to decompose $v$ into $v_i$.
One should note that if $v$ is on the plane of toric diagram
$m[v]$ gives a unit flow.
Again there are ambiguities in $m[v]$ associated with the choice of
perfect matching $m[v_i]$ for each $v_i$ and in 
the way of decomposing $v$ into $v_i$.
The difference
of two different choices of $m[v]$, however, is always a
conserved flow with vanishing height change,
and this ambiguity does not matter in the following arguments.

\paragraph{Shrinking cycles and NS5-branes}
By the T-duality along $\theta_1$ and $\theta_2$, edges
of the base manifold $\B$, or, equivalently,
external lines in the web-diagram are transformed into
NS5-branes wrapped on cycles in the dual $\bT^2$.
The winding number of each NS5-brane is determined in the following
way.
Let us focus on the edge of $\B$ shared by
adjacent two facets $F_\alpha$ and $F_{\alpha+1}$.
This edge is given as the set of points satisfying
\begin{equation}
v_\alpha\cdot w=0,\quad
v_{\alpha+1}\cdot w=0.
\end{equation}
By taking the difference of these two conditions,
we obtain
\begin{equation}
(\Delta p,\Delta q)\cdot (w_1,w_2)=0,
\end{equation}
where $\Delta p=p_{\alpha+1}-p_\alpha$,
$\Delta q=q_{\alpha+1}-q_\alpha$, and $w_1$ and $w_2$ are the first two
components of the vector $w$.
Namely, the external line in the web-diagram
is a line perpendicular to the
side between the corners $v_\alpha$ and $v_{\alpha+1}$ of the toric diagram.

Let $(\theta_1,\theta_2)$ be the coordinate of
$\bT^2$ fiber on the facet $F_\alpha. $\footnote{We have chosen this notation for simplicity, although strictly speaking we should write $(\theta_1^{\alpha},\theta_2^{\alpha})$ since these coordinates depend to $\alpha$.}
These coordinates are inert under the isometry $U(1)[v_\alpha]$.
In order to examine the behavior of the $\bT^2$ fiber near
the external line shared by $F_\alpha$ and $F_{\alpha+1}$,
let us consider the action of $U(1)[v_{\alpha+1}]$ on
the $\bT^2$.
The Killing vector $v_{\alpha+1}$ for the adjacent facet $F_{\alpha+1}$
acts on the coordinate $(\theta_1,\theta_2)$
as
\begin{equation}
\theta_1\rightarrow\theta_1+\Delta p \ t,\quad
\theta_2\rightarrow\theta_2+\Delta q \ t,
\end{equation}
with a parameter $t$.
The corresponding cycle, $(\Delta p,\Delta q)$-cycle,
shrinks at the boundary between
$F_\alpha$ and $F_{\alpha+1}$.
By the T-duality transformation along $\theta_1$ and $\theta_2$,
this shrinking cycle
is transformed to NS5-brane wrapped on $(\Delta q,-\Delta p)$-cycle.
In the bipartite graph, these
cycles wrapped by the NS5-branes are represented as the
zig-zag paths.
(See Figure \ref{c320.eps} (c). There cycles of NS5-branes are denoted by $\mu,\nu,\rho$.)

\begin{figure}[htb]
\centerline{\scalebox{0.5}{\includegraphics{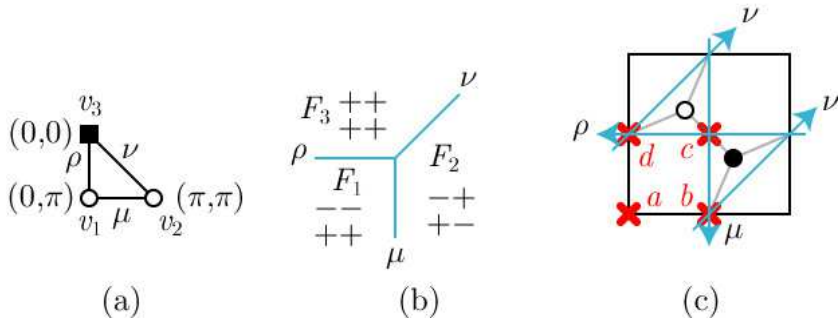}}}
\caption{Some diagrams for $\bC^3$ orientifold are shown.
(a) shows the toric diagram.
The square box shows the choice of $(2\bZ)^2$ sublattice,
and the number placed at each vertex represents
shift variable $(\pi s_1,\pi s_2)$ on the facet
corresponding to each vertex. (b) shows the web-diagram,
together with the assignment of RR-charges on each facet.
Note RR-charges change when crossing cycles of NS5-branes,
just as in the case of conifold
shown in Figure \ref{conifold2.eps}.
(c) shows the bipartite graph, and $a,b,c,d$ denotes
the position of O5-planes.}
\label{c32.eps}
\end{figure}

\begin{figure}[htb]
\centerline{\scalebox{0.5}{\includegraphics{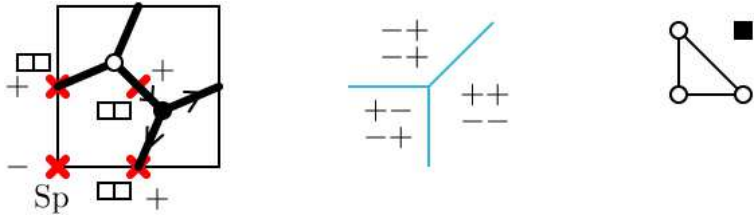}}}
\caption{Another example of orientifold of $\bC^3$ is shown.
See also Figure \ref{c32.eps}.
In this case the square vertex is not contained
in the lattice points of the toric diagram,
and the parity flow
is obtained as a linear combination of perfect matchings.}
\label{c3.eps}
\end{figure}

%%%%%%%%%%%%%%%%%%%%%%%%%%%%%%%%%%%%%%%%%%%%%%%%%%%%%%%%%%%%%%%
\paragraph{Orientifold}
Let us consider orientifold ${\cal M}/\bZ_2$.
We only consider $\bZ_2$ which is a subgroup of
the $U(1)^3$ isometry of ${\cal M}$.
As we see below
this type of $\bZ_2$ gives orientifolded fivebrane systems with
O5-planes, which we are interested in.
We can specify the $\bZ_2$ action by specifying
a Killing vector $V$ for $U(1)$ symmetry which include the
$\bZ_2$ as its subgroup.
Without loss of generality we assume that $V$ is a
primitive vector in the integral lattice.
The generator of $\bZ_2$ is given by
\begin{equation}
(\phi_1,\phi_2,\phi_3)
\rightarrow(\phi_1,\phi_2,\phi_3)+\pi V.
\label{vshift}
\end{equation}
Two vectors $V$ and $V'$ with components
different by even integers define the same $\bZ_2$.
Since $\phi_3$ is identified with
the argument of the coordinate $x^8+ix^9$,
third component of $V$ is $1$ (modulo 2).
This means that the vector $V$ can be represented as
a point on the toric diagram.
In order to define the orientifold we only need to
specify the first two components of $V$
mod $2$.
This is graphically represented in toric diagram
by choosing $(2\bZ)^2$ lattice.
When we draw the toric diagram of an orientifolded toric Calabi-Yau,
we will use squares to represent
points in the $(2\bZ)^2$ lattice
while other points are represented as circles.
An example of such a toric diagram for an orientifold of
$\bC^3$ is given in Figure \ref{c32.eps} (a).
We have four choices of sublattice $(2\bZ)^2$,
but in this example, three of them are
equal up to $SL(2,\bZ)$ transformation, so only one of them is shown.
The other possible choice of $(2\bZ)^2$ will be shown later in Figure \ref{c3.eps}.

From the $\bZ_2$ action on $\phi_i$ given in (\ref{vshift}), together with definition of $\theta$ in (\ref{thetadef}), 
we can easily obtain the following action on the $\bT^2$ fiber
on each facet as
\begin{equation}
(\theta_1,\theta_2)\rightarrow(\theta_1-\pi s_1,\theta_2-\pi s_2),
\end{equation}
with the shift variables $s_1$ and $s_2$ on each facet given by 
\begin{equation}
s_\alpha=v_\alpha-V.
\label{ocu3}
\end{equation}

Note that the shift depends on facets $F_\alpha$, and
this formula can be used to determine the shift
for each facet.
In the example in Figure \ref{c32.eps} (a),
the shift of the $\bT^2$ fiber in each facet is given
beside the corresponding vertex.

\paragraph{T-duality of orientifolds}
In order to relate a toric Calabi-Yau cone to a fivebrane system,
we define Cartesian-like coordinate as follows.
Let $x^4$ and $x^6$ be coordinates
parameterizing $\partial \B$, the plane of web-diagram,
and $\rho\geq0$ be the ``distance'' from the boundary $\partial \B$.
We do not need precise form of these coordinates because
we are only interested in the topological structure.
We combine $\rho$ and $\phi_3$ to define $x^8$ and $x^9$ by
$x^8+ix^9=\rho e^{i\phi_3}$.
The coordinates $x^4$, $x^6$, $x^8$, and $x^9$ defined above
are identified with the same coordinates in the fivebrane system,
while $\theta_1$ and $\theta_2$ are the dual coordinates
to the compact coordinates $x^5$ and $x^7$.
The $\bZ_2$ action on these coordinates
is given by
\begin{equation}
(x^4,\theta_1,x^6,\theta_2,x^8,x^9)
\rightarrow
(x^4,\theta_1-\pi s_1,x^6,\theta_2-\pi s_2,-x^8,-x^9).
\end{equation}
If both $s_1$ and $s_2$ are even integers,
there is a codimension-$2$ fixed plane, O7-plane.
Otherwise, there is no fixed plane.

By the T-duality along the compact coordinates $\theta_1$ and $\theta_2$,
the orientifold is transformed to another orientifold
with the geometric $\bZ_2$ action
\begin{equation}
(x^4,x^5,x^6,x^7,x^8,x^9)
\rightarrow
(x^4,-x^5,x^6,-x^7,-x^8,-x^9).
\label{o5action}
\end{equation}
This is the orientifold of the fivebrane systems which
we discuss in this paper.
This orientifold has four O5-planes at
$(x^5,x^7)=(0,0)$, $(\pi,0)$, $(0,\pi)$, and $(\pi,\pi)$.
There is no dependence on $(s_1,s_2)$ in the geometric $\bZ_2$ action
(\ref{o5action}).

The information of $s_1$ and $s_2$ is encoded in the RR-charges of
four O5-planes.
If both $s_1$ and $s_2$ are even integers, there is an O7-plane
on the Calabi-Yau side, and
the dual configuration contains
four O5-planes with the same sign of RR-charge as the O7-plane.
Otherwise, we have no O7-plane on the Calabi-Yau side,
and two O5$^+$ and two O5$^-$ in the fivebrane system
at the position depending on $(s_1,s_2)$\cite{Witten:1997bs}.
The relation between $(s_1,s_2)$ and the charges of O5-planes are
summarized below.
\begin{eqnarray}
(s_1,s_2)=(0,0)&:&\oplane++++ \mbox{ or }\oplane----,\label{ee}\\
(s_1,s_2)=(1,0)&:&\oplane+--+ \mbox{ or }\oplane-++-,\label{oe}\\
(s_1,s_2)=(1,1)&:&\oplane-+-+ \mbox{ or }\oplane+-+-,\label{oo}\\
(s_1,s_2)=(0,1)&:&\oplane--++ \mbox{ or }\oplane++--.\label{eo}
\end{eqnarray}
By the T-duality relations (\ref{ee}-\ref{eo})
we define the map $\sigma$ from charge assignments
$Q$ to $(\bZ_2)^2$ valued vector $s$,
\begin{equation}
\sigma: Q\rightarrow s
\label{ocu4}
\end{equation}

If we use the notation of (\ref{rule1}), then explicit expression for $s=\sigma(Q)$ is given by
\begin{equation}
(-1)^{s_1}=\int_{\balpha} Q,\quad
(-1)^{s_2}=\int_{\bbeta} Q. 
\label{-1sQ}
\end{equation}
Here $\balpha$ (resp. $\bbeta$) denotes $\balpha$-cycle (resp. $\bbeta$-cycle) of $\bT^2$ which passes through two O5-planes. We have two such $\balpha$-cycles (resp. $\bbeta$-cycles), but both of these two gives the same answer, because the product of all four charges is always $+1$.
The relations in (\ref{-1sQ}) show that
we can regard $(s_1,s_2)$ as the relative charges among four RR charges.

For two charge assignments $Q_1=\oplaneone{q_1}$ and $Q_2=\oplaneone{q_2}$, we define their product by componentwise multiplication:
\begin{equation}
Q_1\cdot Q_2 =\oplanetwo{q_1}{q_2}.
\end{equation}
Then you can directly verify the formula
\begin{equation}
\sigma(Q_1\cdot Q_2)=\sigma(Q_1)+\sigma(Q_2) \mod2 .
\label{sigmaprod}
\end{equation}

If we are given the geometric action of $\bZ_2$ on
a Calabi-Yau cone,
we obtain $(s_1,s_2)$, which can be regarded as relative charges of O5-planes,
on each facet
by the relation (\ref{ocu3}).
Conversely, given relative charges, we always have two possible charge assignments as listed in (\ref{ee})-(\ref{eo}), which are related to each other by total charge flip.
We cannot choose one of them only with the information of
the geometric action of $\bZ_2$.

Instead of starting from the geometric $\bZ_2$ action on the Calabi-Yau,
let us assume that we are given
a brane tiling with fixed points specified,
and that we know charge distribution on one facet.
With this information, we can reconstruct all the information about
fivebrane system and $\bZ_2$ action on the Calabi-Yau cone
in the following way.

We first reconstruct the
toric diagram and the web-diagram by using zig-zag paths.
Let $F_\alpha$ and $F_{\alpha+1}$ be two adjacent facets.
If the side between two corners $v_\alpha$ and $v_{\alpha+1}$
includes $n$ edges, there are $n$ parallel zig-zag paths corresponding to the
external lines between facets $F_\alpha$ and $F_{\alpha+1}$.
Let $Z_{\alpha+1,\alpha}$ be the union of these $n$ zig-zag paths.

If the charge assignment $Q_\alpha=\oplane{q_{\alpha1}}{q_{\alpha2}}{q_{\alpha3}}{q_{\alpha4}}$ in the facet $F_\alpha$ is given,
we can determine $Q_{\alpha+1}$ in the next facet $F_{\alpha+1}$ by
flipping the RR-charges of fixed points which are passed through by
$Z_{\alpha+1,\alpha}$.

This relation between $Q_\alpha$ and $Q_{\alpha+1}$ is expressed as
\begin{equation}
Q_{\alpha+1}=Q_{\alpha} \cdot \rho(Z_{\alpha+1,\alpha}). \label{ocu2}
\end{equation}
Here $\rho(Z_{\alpha+1,\alpha})$ is the charge assignment such that if
a fixed point is on a zig-zag path in $Z_{\alpha+1,\alpha}$,
the charge of the fixed point is $-1$ while the charge is $1$ otherwise.

Let us take the orientifold of $\bC^3$
shown in Figure \ref{c32.eps} as an example.
By using the bipartite graph (c), we can 
draw the corresponding toric diagram and web-diagram.
The toric diagram has three corners and correspondingly
there are three facets in the web-diagram.
We assume that we know the charge distribution
on one of the facets, say, $F_1$.
The charge assignments on other facets are determined as follows.
If we move from $F_1$ to $F_2$ on the web-diagram, we cross the NS5-brane $\mu$.
The Figure \ref{c32.eps} (c)  shows that
this NS5-brane intersects with O5-plane $b$ and $c$.
and we can obtain the charge assignment on $F_2$ from that on $F_1$ by
flipping the RR-charges of $b$ and $c$. More formally, we have
$Q_1=\oplane{+}{+}{-}{-}$,
$Q_2=\oplane{+}{-}{+}{-}$, and
$\rho(Z_{2,1})=\oplane{+}{-}{-}{+}$,
and this satisfies the relation (\ref{ocu2}).
By repeating this procedure, we obtain charge assignments on all the facets.

The relation (\ref{ocu2}) is consistent with the relation (\ref{ocu3}). In order to see this, we use the following relation, which is proved in Appendix \ref{eqns.sec}:
\begin{equation}
\sigma(\rho(Z_{\alpha+1,\alpha}))=h(Z_{\alpha+1,\alpha})
\mod2 .
\label{ocu5}
\end{equation}
By using (\ref{mviab}), (\ref{sigmaprod}), and (\ref{ocu5})
the relation (\ref{ocu2}) is mapped by $\sigma$ to
\begin{equation}
s_{\alpha+1}-s_\alpha=v_{\alpha+1}-v_\alpha,
\label{ocu6}
\end{equation}
and this shows that
the relation (\ref{ocu2}) is ``integrable'', and consistently determine the vector $V$ modulo $2$.

\paragraph{$\bZ_2$ parity of mesonic operators}
Mesonic operators in a gauge theory are represented as
closed paths in the brane tiling.
Let $C[{\cal O}]$ be the closed path corresponding to a mesonic operator
${\cal O}$.
Mesonic operators are important
when we relate gauge theories and toric Calabi-Yau cones
because we can regard mesonic operators as holomorphic monomial
functions in the Calabi-Yau cone.
In other words, we can use (an appropriate subset of) mesonic operators as coordinates
in the Calabi-Yau.
To establish the relation between mesonic operators and
monomial functions in the toric Calabi-Yau, we can use
charges associated with $U(1)^3$ symmetry.
For both mesonic operators and monomial functions we can
assign three charges, and by these charges
we can establish one-to-one correspondence
between mesonic operators and monomial functions.

As we mentioned above, we can specify $U(1)$ symmetry by a vector $v\in\Gamma$.
We can determine the charges of mesonic operators for a given $U(1)$ by
the following relation
\begin{equation}
\mbox{$U(1)[v]$ charge of an operator ${\cal O}$} = \langle m[v],C[{\cal O}]\rangle,
\label{U1charge}
\end{equation}
where $\langle f,C\rangle$ is the flux of $f$ across $C$ defined
above
(\ref{heightdef}).
Because the path for a gauge invariant mesonic operator is closed,
the ambiguity in the definition of $m[v]$ does not
affect the flux.

As a special case of this relation, the $\bZ_2$ parity of a mesonic operator ${\cal O}$
under the transformation (\ref{vshift}) can be obtained
from the $U(1)[V]$ charge of the mesonic operator.
If the charge is even (odd), the parity is $+$ ($-$).
In other words, we can determine the $\bZ_2$ parity
$P[{\cal O}]$ for an operator ${\cal O}$
by the mod $2$ flux
of the unit flow $m[V]$
across $C[{\cal O}]$.
\begin{equation}
P[{\cal O}]=(-1)^{\langle m[V],C[{\cal O}]\rangle} .
\label{ocu10}
\end{equation}
We refer to the unit flow $m[V]$ as a parity flow.

If the toric diagram includes square vertices,
we can use one of perfect matchings associated with
one of square vertices in the toric diagram as
a parity flow.
Otherwise, we have to use linear combination of
perfect matchings.
One such example is shown in Figure \ref{c3.eps}.

%%%%%%%%%%%%%%%%%%%%%%%%%%%%%
\subsection{T-parity and RR-charge}
The purpose of this subsection is
to clarify the relation
between the T-parity and the RR-charge by comparing
the formula (\ref{ocu10}) for $\bZ_2$ parity
obtained in the previous subsection
and the rules proposed in \cite{Franco:2007ii}.

\paragraph{Superpotential rule}
We can easily reproduce the superpotential rule (Rule \ref{superpotential.rule})
from the formula (\ref{ocu10}).
Because terms in the superpotential correspond to
cycles enclosing only one vertex in a bipartite graph,
the rule is equivalent to
the statement that
the $\bZ_2$ parity of a mesonic operator ${\cal O}$ is
given by the flux of 
an odd flow, flow with odd source and sink at each vertex,
across the path $C[{\cal O}]$.
Because $m[V]$ is a unit flow
it correctly gives the $\bZ_2$ parity satisfying the superpotential rule.

\paragraph{Product rule}
Next we are going to reproduce the product rule (Rule \ref{product.rule}). Let us define $\rho(m_{\alpha})$
as the charge assignment such that if a fixed point
is on the perfect matching $m_{\alpha}$
the charge of the fixed point is $-1$,
while the charge is $+1$ otherwise.
This definition is similar to that of $\rho(Z_{\alpha+1,\alpha})$,
and we have the following relation:
\begin{equation}
\rho(Z_{\alpha+1,\alpha})=\rho(m_{\alpha+1})\cdot\rho(m_{\alpha}).\label{Zmm}
\end{equation}

In order to understand this relation, recall (\ref{mviab}) says
\begin{equation}
m_{\alpha+1}-m_\alpha=Z_{\alpha+1,\alpha} \mod \textrm{(boundaries)},
\label{ocu7}
\end{equation}
since winding numbers,
or height function of both sides of this equation coincides.
By definition, this almost proves (\ref{Zmm}).
The only remaining problem is the possible contribution
to $\rho$ from boundary terms,
which are conserved flows with vanishing height change.
Namely, boundaries might pass through fixed points and contribute to $\rho$.
It turns out, however, that this contribution is absent
since it is impossible for boundaries to pass through fixed points
and be $\bZ_2$-symmetric at the same time. This proves (\ref{Zmm}).

Now if we use (\ref{Zmm}), we can rewrite the relation (\ref{ocu2}),
which  represents the RR-charge flip of O5-planes at intersections with NS5-branes, into the following form:
\begin{equation}
Q_{\alpha+1}\cdot \rho(m_{\alpha+1})=Q_\alpha \cdot \rho(m_\alpha).
\label{ocu8}
\end{equation}
This relation means that $Q_\alpha \cdot \rho(m_\alpha)$ in fact does not
depend on facets,
and it is possible to define facet-independent charge $T'$ by
\begin{equation}
T'=Q_\alpha \cdot \rho(m_\alpha).
\label{ocu9}
\end{equation}
We propose the charge assignment $T'$ defined by
this equation is nothing but the T-parity $T$.

For this charge $T'$ to be acceptable as the T-parity,
we should confirm that $T'$ defined in (\ref{ocu9})
combined with the formula (\ref{rule1}) gives the
same $\bZ_2$ parity as (\ref{ocu10}) for
mesonic operators which are described by $\bZ_2$ symmetric cycles.
This can be easily shown by
rewriting the formula (\ref{ocu10})
in terms of $T'$.
We first use (\ref{ocu3}) and rewrite (\ref{ocu10}) as
\begin{equation}
P[{\cal O}]
=(-1)^{\langle m_\alpha,C[{\cal O}]\rangle}
(-1)^{\langle m[s_\alpha],C[{\cal O}]\rangle} .
\label{ocu11}
\end{equation}
In the Appendix \ref{eqns.sec}, we prove the following two formulae:
\begin{equation}
(-1)^{\langle m_\alpha,C[{\cal O}]\rangle}
=\int_{C[{\cal O}]}\rho(m_\alpha),
\label{ocu12}
\end{equation}
and
\begin{equation}
(-1)^{\langle m[s_\alpha],C[{\cal O}]\rangle}
=\int_{C[{\cal O}]}Q_\alpha,
\label{ocu13}
\end{equation}
where $\int_{C[{\cal O}]}$ means the product of charges of fixed points
passed through by the path $C[{\cal O}]$, as defined in (\ref{rule1}).
The first says that if $m_\alpha$ is the perfect matching for $\alpha$,
which is $\bZ_2$ symmetric,
and if $C[{\cal O}]$ is a $\bZ_2$ symmetric path,
only the fixed points contribute to the mod $2$ intersection number.
Substituting (\ref{ocu12}) and (\ref{ocu13}) into (\ref{ocu11})
we immediately obtain
\begin{equation}
P[{\cal O}]=\int_{C[{\cal O}]}T',
\label{ocu14}
\end{equation}
and this is nothing but the formula (\ref{rule1}) with $T$ replaced by $T'$.

Note that not only $T=T'$ but also $T=-T'$ give the same $\bZ_2$
parity of mesonic operators, and we still have the following two possibilities:
\begin{equation}
T=\pm T'.
\end{equation}

In order to determine the overall sign of the T-parity,
we need additional information, or assumption
about the relation between overall sign of
RR-charge and overall sign of T-parity.

The reason why we cannot simply identify
RR-charges and T-parities is that
as we mentioned above 
the RR-charge may change depending on the facets.
This is, however, only the case for the O5-planes
on edges.
If an O5-plane is inside a face,
its RR-charge is everywhere the same,
and we can simply identify the RR-charge as T-parity.
So, let us adopt the following assumption:
\begin{itemize}
\item
For fixed points on faces, the T-parity is the same
as the RR-charge of the O5-plane.
\end{itemize}
If there exist fixed points on faces,
this condition uniquely determine the
T-parity as
\begin{equation}
T=T'=Q_\alpha\cdot \rho(m_\alpha).
\label{tandrr}
\end{equation}

There is a simple relation
among RR-charge and T'-charge of O5-planes.
The definition (\ref{ocu9}) of $T'$
can in fact be rewritten in the following form which does not
directly refer to perfect matchings:
\begin{Rule}[Angle rule]
\item
When O5-plane consists of two parts with
opposite RR-charges,
the $T'$ charge of the O5-plane
is the same as the RR-charge of the O5-plane
occupying the major angle.
\label{angle.rule}
\end{Rule}
This can be shown from (\ref{ocu9}) and
the following theorem proved in the Appendix \ref{thm.sec}:
\begin{theorem}\label{minor}
Let $I$ be an edge in a bipartite graph,
and $\{F_\alpha,F_\beta,\ldots,F_\gamma\}$ be the set of
facets whose associated perfect matchings include the edge $I$.
Then, facets in the set
$\{F_\alpha,F_\beta,\ldots,F_\gamma\}$ form
one continuous region
in the web-diagram,
and the central angle of the region is always a minor angle.
\end{theorem}

The angle rule (Rule \ref{angle.rule}) shows that the $T'$-charge
of an O5-plane is determined
by the local information about the O5-plane
without using charges of other orientifold planes.
If we assume this is the case for the T-parity,
it is natural that
the T-parity for a fixed point is always determined by
(\ref{tandrr}) regardless of the existence of
O5-planes inside faces.

In the next section, we give
another reason
why we should choose the overall sign of T-parity
as the equation (\ref{tandrr}).

%%%%%%%%%%%%%%%%%%%%%%%%%%%%%%%%%%%%%%%%%%%%%%%%%%%%%%%%%
\section{Flavor branes}\label{d7.sec}
%%%%%%%%%%%%%%%%%%%%%%%%%%%%%%%%%%%%%%%%%%%%%%%%%%%%%%%%%
\subsection{Quark mass terms}\label{qmt.sec}
In the previous section we discussed only the mesonic operators
made of bi-fundamental fields.
Let us turn to quark fields in the (anti-)fundamental representation,
which emerge when we introduce flavor branes.

We here only discuss flavor D5-branes parallel to the
O5-planes in Table \ref{branes.tbl}.
In this subsection we do not consider orientifolds.
By T-duality transformation
they are transformed into D7-branes wrapped on divisors in the
toric Calabi-Yau geometry.
In \cite{Franco:2006es} graphical representation of such flavor D7-branes
and corresponding superpotential terms are proposed.
A D7-brane wrapped on a divisor in the Calabi-Yau $3$-fold is
represented as a curve connecting two punctures
on the NS5-brane worldvolume,
which is referred to as Riemann surface in \cite{Franco:2006es}.
In fivebrane diagrams, these two punctures
are represented as two cycles,
and the curve connecting two punctures corresponding to an
intersection of these cycles.
This intersection point is nothing but the flavor D5-brane worldvolume
projected onto the 57-plane.

\begin{figure}[htb]
\centerline{\scalebox{0.5}{\includegraphics{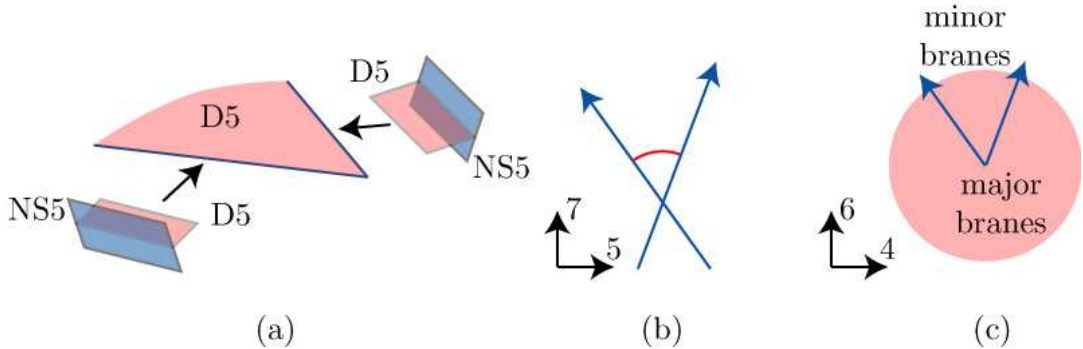}}}
\caption{(a) shows a flavor brane stretched between two legs
in a web-diagram.
In the internal 57-space, the D5-brane is attached to the NS5-branes.
In the corresponding fivebrane diagram (b) we use an arc to represent the flavor brane.
As is shown in (c) there are two possible flavor branes
associated with a pair of legs in the web-diagram.
We call them minor and major branes.
}
\label{flavor.eps}
\end{figure}

On the web-diagram, a flavor brane is represented as
a fan between two external legs corresponding to the
two zig-zag paths
((a) in Figure \ref{flavor.eps}).
%%%%%%
When we specify two legs, there are always two
fans defined by these legs.
One has minor central angle and the other has
major central angle.
We call corresponding two possible flavor branes
for a given pair of external legs
minor branes and major branes
((c) in Figure \ref{flavor.eps}).
In order to distinguish these two types of flavor branes in fivebrane systems,
we represent flavor branes as arcs at the intersection of cycles
((b) in Figure \ref{flavor.eps}).
Arcs are drawn in the angles corresponding to the fans
on the web-diagram.
We can define these angles because the directions of the cycles
are the same as the directions of external legs in the web-diagram.

\paragraph{Minor flavor branes}
In \cite{Franco:2006es}
the following superpotential
is proposed for quarks $q$ and $\wt q$ emerging by
the introduction of flavor branes placed on an intersection $I$:
\begin{equation}
W=\wt q\Phi_Iq,
\label{QPQ}
\end{equation}
where $\Phi_I$ is the bi-fundamental field
associated with the intersection $I$.
This superpotential corresponds only to
minor flavor branes as
will be confirmed in the following.

If we assume that quark fields are supplied from
D3-D7 strings in the Calabi-Yau perspective, the fundamental fields must become
massless when D3-branes coincide with the D7-branes.
Namely, massless loci of quark fields in the moduli space
should be identified with the worldvolume of the D7-branes.
(In this paper we consider only the Coulomb branch,
in which quarks have vanishing vevs.)

When the quark mass term is given by (\ref{QPQ}),
the massless locus is given by $\Phi_I=0$.
(Following the usual procedure to obtain Calabi-Yau geometry,
we here treat all the gauge groups as $U(1)$).
In order to determine the corresponding divisor in the moduli space,
we should solve the F-term conditions imposed on bi-fundamental fields.
The solution is given by
\cite{Franco:2006gc}
\begin{equation}
\Phi_I=\prod_{\alpha'\ni I}\rho_{\alpha'},
\label{ppr}
\end{equation}
where $\rho_{\alpha'}$ are complex fields
defined for each perfect matching $\alpha'$,
and $\alpha'\ni I$ means that the product is taken over all
the perfect matchings which include the edge $I$.
By this relation we can describe the moduli space
of quiver gauge theory as the moduli space of
gauged linear sigma model (GLSM) with the fields $\rho_{\alpha'}$.
The equation (\ref{ppr}) means that
the massless locus is given by the union
of loci defined by $\rho_{\alpha'}=0$.
Because we are interested in divisors,
we do not take care of subspace of moduli space with dimension
less than $2$.
We only focus on the submanifold ${\cal M}'\subset{\cal M}$
which is defined as the complement of the submanifold
corresponding to the legs and the center of the
web-diagram.
We can show that
in this submanifold
GLSM fields $\rho_{\alpha'}$ which do not correspond to
corners of the toric diagram do not vanish.
This allow us to forget about such fields and we have only to
take care of fields $\rho_\alpha$, which correspond to corners in
the toric diagram.
The following theorem can be proved:
\begin{theorem}\label{divisor}
In the subspace ${\cal M}'$,
the divisor corresponding to a
facet $F_\alpha$ is given by $\rho_\alpha=0$ in the GLSM.
\end{theorem}
The proof is given in Appendix \ref{GLSM.sec}.
With this theorem we obtain
\begin{equation}
\mbox{massless locus}=\bigcup_{\alpha\ni I}F_{\alpha},
\end{equation}
where we use $F_\alpha$ for the divisor corresponding to the facet.
The theorem \ref{minor} means that this is nothing but the
worldvolume of the minor branes associated with the edge $I$.

\begin{figure}[htb]
\centerline{\scalebox{0.5}{\includegraphics{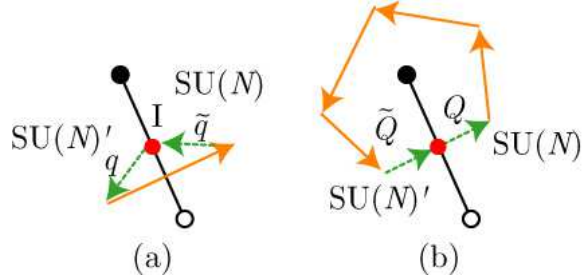}}}
\caption{Closed paths representing quark mass terms
for minor flavor branes (a) and major flavor branes (b) are shown.}
\label{fundmass0.eps}
\end{figure}

\paragraph{Major flavor branes}
In order to obtain the worldvolume of major flavor branes,
we need different quark mass terms from (\ref{QPQ}).
Let us assume the following form of quark mass terms:
\begin{equation}
W=\wt Q{\cal O}Q,
\label{WQOQ}
\end{equation}
where ${\cal O}$ is composite operator made of bi-fundamental fields.
We denote quark fields provided by major flavor branes by $Q$ and $\wt Q$
while we write $q$ and $\wt q$ the quark fields for minor branes.
From the theorem \ref{minor},
the worldvolume of major flavor branes associated with the
edge $I$ is given by
\begin{equation}
\mbox{major branes}=\bigcup_{\alpha\notni I}F_{\alpha}.
\end{equation}
where $\alpha\notni I$ means that the product is taken over all
the perfect matchings which do not include the edge $I$
and are associated with corners of the toric-diagram.
By the theorem \ref{divisor}
this is given by ${\cal O}=0$ with the operator ${\cal O}$ defined by
\begin{equation}
{\cal O}=\prod_{\alpha'\notni I}\rho_{\alpha'}.
\end{equation}
In order to write the superpotential (\ref{WQOQ}),
we need to rewrite the operator ${\cal O}$
in terms of bi-fundamental fields in the gauge theory.
It is easy to see that
\begin{equation}
{\cal O}=\prod_{J\in k, J\neq I}\Phi_J,
\label{opp}
\end{equation}
where $k$ is one of two endpoints of the edge $I$,
and $J\in k$ means edges sharing the vertex $k$ as their endpoints.
When we regard this as the operator in the gauge theory with
non-Abelian gauge group the constituent fields should
be ordered so that the color indices
of adjacent fields match.
The operator ${\cal O}$ is graphically represented as the
path consisting of solid arrows in Figure \ref{fundmass0.eps} (b).
In the definition of the operator ${\cal O}$
there are two choices of the endpoint of the edge $I$.
Let ${\cal O}_B$ and ${\cal O}_W$ be the two operators
obtained by choosing black and white endpoints of $I$, respectively.
Because the superpotential of bi-fundamental fields
includes
\begin{equation}
W=\tr(\Phi_I{\cal O}_B)-\tr(\Phi_I{\cal O}_W),
\end{equation}
and the F-term condition of $\Phi_I$ gives
\begin{equation}
{\cal O}_B={\cal O}_W,
\end{equation}
the superpotential (\ref{WQOQ}) does not depend on the
choice between ${\cal O}_B$ and ${\cal O}_W$.

\begin{figure}[htb]
\centerline{\scalebox{0.5}{\includegraphics{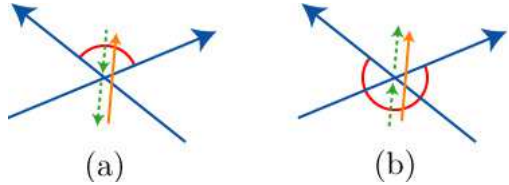}}}
\caption{Graphical representation of minor branes (a) and major branes (b),
and corresponding fields are shown.}
\label{minormajor.eps}
\end{figure}

Let us compare the superpotentials
(\ref{QPQ}) for minor branes and (\ref{WQOQ}) for major branes.
\begin{equation}
W_{\rm minor}=\wt q_i\Phi_I^i{}_{j'}q^{j'},\quad
W_{\rm major}=\wt Q_{i'}{\cal O}^{i'}{}_jQ^j.
\label{wmwm}
\end{equation}
These two are represented as cycles made of dashed and solid arrows
in Figure \ref{fundmass0.eps}.
In (\ref{wmwm}) color indices are explicitly written.
Notice that the existence of these terms requires
the chirality of the quark fields should be opposite
between minor and major flavor branes.
If we have a bi-fundamental field in the representation $(\fund,\fundbar)$
at the edge $I$,
minor branes give quarks in the representation
$(\fundbar,1)$ and $(1,\fund)$ while
major branes give ones in
$(\fund,1)$ and $(1,\fundbar)$
(Table \ref{quarks}).
\begin{table}[htb]
\caption{The representations of quark fields for
two types of flavor branes are shown.}
\label{quarks}
\begin{center}
\begin{tabular}{cl}
\hline
\hline
& $SU(N)\times SU(N)'$ \\
\hline
bi-fund. & $\Phi_I^i{}_{j'}(\fund,\fundbar)$ \\
minor brane & $q^{i'}(1,\fund)$, $\wt q_i(\fundbar,1)$ \\
major brane & $Q^i(\fund,1)$, $\wt Q_{i'}(1,\fundbar)$ \\
\hline
\end{tabular}
\end{center}
\end{table}
This difference is important when we use flavor branes
to cancel the gauge anomaly associated with orientifold planes.
In Figure \ref{minormajor.eps}, the difference of quark representations
is expressed by the orientation of dashed arrows.

%%%%%%%%%%%%%%%%%%%%%%%%%%%%%%%%%%%%%%%%%%%%%%%%%%%%%%
\subsection{Orientifold planes and flavor branes}
As we mentioned in \S\ref{anom.sec},
if an O5-plane is divided into
major and minor O5-planes
these two parts carry opposite RR-charges to each other.
One way to compensate the charge flip of O5-planes
is to introduce appropriate number of flavor branes coinciding
with the O5-planes.
Another way to realize the charge conservation using
RR-charge flow along NS5-branes is
discussed in the next subsection.

Let us first consider the case with positive T-parity.
In this case, by the angle rule (Rule \ref{angle.rule})
with $T'$ identified with $T$, the minor and major O5-planes
carry negative and positive RR-charges, respectively.
We can match the RR-charges of minor and major parts
by introducing $k+4$ minor branes and $k$ major branes.
In the parent theory these flavor branes provide
quarks shown in Table \ref{quarks}.
By the orientifold projection,
the two gauge groups $SU(N)$ and $SU(N)'$ on the both sides
of the edge $I$ are identified,
and the bi-fundamental field $\Phi_I^i{}_{j'}$
in $(\fund,\fundbar)$ representation
becomes field $\Phi_I^{\{ij\}}$ in the symmetric representation $\symm$
of $SU(N)$.
The quark fields $q$ and $\wt q$, and $Q$ and $\wt Q$ are
identified, and independent fields are
$q$ in $\fundbar$ and $Q$ in $\fund$.
(Table \ref{tplus.tbl})
\begin{table}[htb]
\caption{Fields arising at the fixed point with positive T-parity are shown.
$SU(N)$ is the gauge group and $SO(k+4)\times Sp(k/2)$ is
the flavor symmetry.}
\label{tplus.tbl}
\begin{center}
\begin{tabular}{cccc}
\hline
\hline
       & $SU(N)$ & $SO(k+4)$ & $Sp(k/2)$ \\
\hline
$\Phi$ &    $\symm$  &  $1$  & $1$  \\
$q$ &    $\fundbar$  &  $\fund$  & $1$  \\
$Q$ &    $\fund$  &  $1$  & $\fund$  \\
\hline
\end{tabular}
\end{center}
\end{table}

The superpotential is given by
\begin{equation}
W_{(T=+)}
=\delta_{AB}q_i^A\Phi^{\{ij\}}q_j^B
+J_{CD}Q^{Ci}{\cal O}_{ij}Q^{Dj},
\end{equation}
where ${\cal O}$ is the composite operator
defined in the parent theory by (\ref{opp}).
The flavor symmetry is $SO(k+4)\times Sp(k/2)$,
and $\delta_{AB}$ and $J_{CD}$ are the symmetric
and antisymmetric invariant tensors of
the flavor groups $SO(k+4)$ and $Sp(k/2)$, respectively.

The fields arising at the fixed point contribute to the
$SU(N)$ gauge anomaly by
\begin{equation}
d_{\symm} + (k+4)d_{\fundbar}+kd_{\fund}=Nd_{\fund},
\end{equation}
and this is the same
as the contribution of the bi-fundamental field in the parent theory.
Therefore, if the parent theory is anomaly free before the introduction
of flavor branes,
the daughter theory is also automatically anomaly free.

If the T-parity is negative,
the minor and major O5-planes
carry positive and negative RR-charges, respectively.
We can match the RR-charges of two parts by introducing
$k$ minor and $k+4$ major flavor branes.
In the parent theory,
in addition to the bi-fundamental field in $(\fund,\fundbar)$,
we have quark fields shown in Table \ref{quarks}.
By the orientifold projection, these fields become
an antisymmetric tensor field $\Phi^{[ij]}$ in $\asymm$,
$q$ in $\fundbar$, and $Q$ in $\fund$.
The flavor symmetry is $SO(k+4)\times Sp(k/2)$.
The gauge and flavor quantum numbers are shown in Table \ref{tminus.tbl}.
\begin{table}[htb]
\caption{Fields arising at the fixed point with negative T-parity are shown.
$SU(N)$ is the gauge group and $SO(k+4)\times Sp(k/2)$ is
the flavor symmetry.}
\label{tminus.tbl}
\begin{center}
\begin{tabular}{cccc}
\hline
\hline
       & $SU(N)$ & $SO(k+4)$ & $Sp(k/2)$ \\
\hline
$\Phi$ &    $\symm$  &  $1$  & $1$  \\
$q$ &    $\fundbar$  &  $1$ & $\fund$  \\
$Q$ &    $\fund$  & $\fund$ & $1$ \\
\hline
\end{tabular}
\end{center}
\end{table}

The superpotential is given by
\begin{equation}
W_{(T=-)}
=\delta^{AB}q_A^i {\cal O}_{ij}q_B^j
+J^{CD}Q_{Ci}\Phi^{[ij]}Q_{Bj}.
\end{equation}

As well as the case of positive T-parity,
the fields arising at the fixed point contribute the $SU(N)$
gauge anomaly by the same amount as the bi-fundamental field in the
parent theory
\begin{equation}
d_{\asymm}
 + (k+4)d_{\fund}
 + k d_{\fundbar}
 =Nd_{\fund},
\end{equation}
and the anomaly cancels if so does the anomaly in the parent theory before introducing the
flavor branes.

Note that in the above we assumed that $T=+T'$ when we use Rule \ref{angle.rule}
to determine the RR-charge of minor and major O5-planes.
If we take the opposite sign $T=-T'$,
the chirality of quark fields arising at the fixed points
are reversed, and the gauge anomaly does not cancel.
This is another reason why we should take $T=+T'$.

%%%%%%%%%%%%%%%%%%%%%%%%%%%%%%%%%%%%%%%%%%%%%%%%%%%%%%%%%%%%%%%%
\subsection{Flow of flavor brane charge}
In the previous subsection
we discuss the relation between RR charge conservation
and gauge anomaly cancellation in the case of orientifold.
We show that if there are appropriate number of flavor branes
coinciding with
O5-planes the gauge anomaly cancels.
The requirement of the existence of
flavor branes coinciding with O5-planes,
however, is too restrictive.
We can loosen this condition by using the flow of the D5-brane charge
along NS5-branes.

Let us first consider un-orientifolded case.
As an example let us suppose that we introduce $N_f$ minor branes
at an edge $I$.
In general we need to introduce more flavor branes to
realize the charge conservation and the gauge anomaly cancellation.
A simple way is to introduce the same number of major
flavor branes at the same edge.
Then the quark spectrum becomes vector-like and the gauge anomaly cancels.
The charge conservation of flavor branes is also obvious.
This is, however, not the unique way to fulfill these consistency conditions.

\begin{figure}[htb]
\centerline{\scalebox{0.5}{\includegraphics{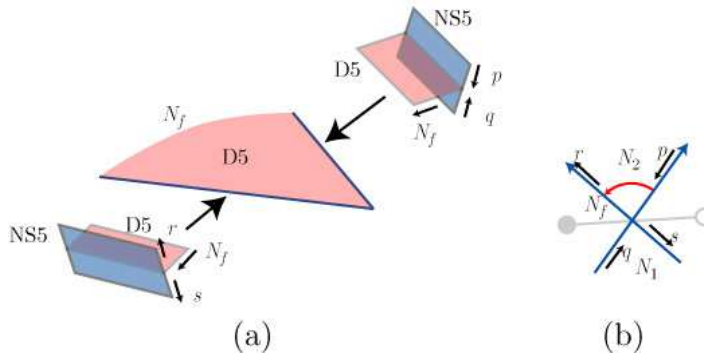}}}
\caption{D5-brane charge carried by one NS5-brane is
transfered into another NS5-brane by flavor D5-branes.
The charge conservation requires $p+q=N_f=r+s$.}
\label{flavorflow.eps}
\end{figure}

Look at Figure \ref{flavorflow.eps}.
It shows $N_f$ minor flavor branes between two external legs in
the web-diagram.
These flavor branes carry D5-brane charge $N_f$,
and the charge is supplied from one of the NS5-branes and
flows into another NS5-brane on the other side.
Even if we do not introduce major flavor branes at the edge $I$,
the D5-charge can flow on NS5-branes, and it can be consistently conserved
if we arrange the flow in the network of NS5-branes appropriately.
In the example shown in Figure \ref{flavorflow.eps},
the following charge conservation condition must hold:
\begin{equation}
p+q=N_f=r+s,
\label{flavorcharge}
\end{equation}
where $p$, $q$, $r$, and $s$ are the D5-charges
flowing on NS5-branes as shown in
Figure \ref{flavorflow.eps}.

The flow of the D5-brane charge on NS5-branes are
graphically described as flows along cycles in fivebrane diagrams.
At intersections, these D5-charges can be transfered from one cycle to
another, and the amount of the transfered charge is
determined by the numbers of the
minor and major flavor branes.
The direction of the charge transfer is
represented as arrows on the arcs in the diagram.
The supersymmetry requires these arrows are always in the same direction
because a flow in the opposite direction means that the corresponding
flavor branes carry negative RR-charge.
We take the convention in which all the arcs are counter-clockwise.

The charge carried by cycles must be taken into account when we
determine the charge of color D5-branes, the numbers assigned to faces
in the bipartite graph.
The number assigned to each face in a fivebrane diagram
must be determined so that
the difference of numbers assigned to adjacent faces
is equal to the flow along the cycle shared by the faces.
See also Figure \ref{inflow.eps}.
In the example of Figure \ref{flavorflow.eps} (b),
the two numbers $N_1$ and $N_2$ must satisfy
\begin{equation}
N_2-N_1=p-s=r-q.
\end{equation}
If the conditions such as (\ref{flavorcharge}) are satisfied
at each intersection,
we can consistently determine the charges of faces.

Generalization to orientifolds is straightforward.
When we consider charge flow along cycles,
we treat a positive O5-plane just like
four coincident flavor branes
because $O5^+$ transfers the charge by $+2$ from
one cycle to another
while $O5^-$ transfers charge $-2$
in the opposite direction.
We should also impose the $\bZ_2$-symmetry on
the flow.
Because the D5-brane charge does not change its sign by the
orientifold flip,
the flow of the D5-brane charge should be invariant under
the rotation of the diagram by angle $\pi$.
See Figure \ref{ex1.eps} for an example.

%%%%%%%%%%%%%%%%%%%%%%%%%%%%%%%%%%%%%%%%%%%%%%%%%%%%%%%%%%%%%%%%%
\subsection{Charge conservation and anomaly cancellation}
In this section we prove that if a fivebrane system
respects the D5-brane charge conservation law
the gauge anomalies of the corresponding quiver gauge theory
cancel.

In order to give a number assignment to faces, cycles, and flavor branes
so that the
D5-brane charge conserves, we should first assign
numbers to faces in fivebrane diagrams.
\begin{figure}[htb]
\centerline{\scalebox{0.5}{\includegraphics{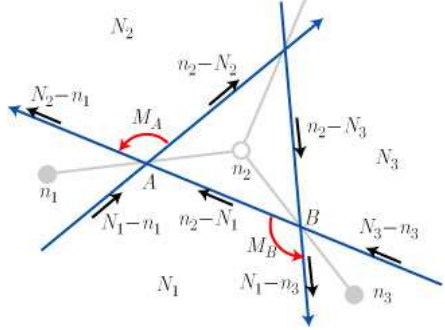}}}
\caption{The flow on cycles and the transfer at
intersections are uniquely determined by the
charge assignment to faces.}
\label{d5charge.eps}
\end{figure}
In a fivebrane diagram, there are two kinds of faces:
faces corresponding to faces in the bipartite graph and
ones corresponding to vertices in the bipartite graph.
We assign numbers to both these two types of faces.
At this step we have no restriction except that
the numbers assigned to faces of the bipartite graph
must be non-negative.

Once we assign numbers to faces in the fivebrane diagram,
the numbers for each segment
in cycles are
uniquely determined by the charge conservation law.
The number assigned to a part of a cycle shared by two faces is given by the
difference of the charges assigned to the two faces.
(Figure \ref{inflow.eps})

Finally, after we obtained all the charges assigned to cycles
in this way, we determine the charge transfer between two intersecting cycles
at every intersection
by the charge conservation relation such as (\ref{flavorcharge}).
For the two intersection $A$ and $B$ in Figure \ref{d5charge.eps},
the flows are
\begin{equation}
M_A=N_1+N_2-n_1-n_2,\quad
M_B=N_1+N_3-n_2-n_3.
\end{equation}
These numbers gives the difference between
numbers of minor and major branes inserted at each
intersection.
\begin{equation}
M_A=N^{\rm minor}_A-N^{\rm major}_A,\quad
M_B=N^{\rm minor}_B-N^{\rm major}_B.
\end{equation}
It may be convenient to draw arcs for both minor and major flavor branes
and assign them the numbers $N^{\rm minor}$ and $N^{\rm major}$
to embed all the information of the brane system in the diagram.
In Figure \ref{d5charge.eps} we show only the net charge transfer
at each intersection.

Let us confirm that the $SU(N_1)$ anomaly
cancels in the example shown in Figure \ref{d5charge.eps}.
The chiral multiplets arising at the intersection $A$
belong to the following representations
of gauge group $SU(N_1)\times SU(N_2)\times SU(N_3)$:
\begin{equation}
(\fundbar,\fund,1)
+N^{\rm minor}_A[(\fund,1,1)+(1,\fundbar,1)]
+N^{\rm major}_A[(\fundbar,1,1)+(1,\fund,1)]
\end{equation}
These contribute the $SU(N_1)$ anomaly in the unit of $d_\fund$ by
\begin{equation}
-N_2+N^{\rm minor}_A-N_A^{\rm major}=
-N_2+M_A=
N_1-n_1-n_2.
\label{anom1}
\end{equation}
At the intersection $B$, we have the following chiral multiplets
\begin{equation}
(\fund,1,\fundbar)
+N^{\rm minor}_B[(\fundbar,1,1)+(1,1,\fund)]
+N^{\rm major}_B[(\fund,1,1)+(1,1,\fundbar)],
\end{equation}
and these contribute to the $SU(N_1)$ anomaly by
\begin{equation}
N_3-N_B^{\rm minor}+N_B^{\rm major}=
N_3-M_B=
-N_1+n_2+n_3.
\label{anom2}
\end{equation}
In the anomalies (\ref{anom1}) and (\ref{anom2}),
the contributions of $N_2$ and $N_3$ have already canceled.
Furthermore,
if we add these two contributions (\ref{anom1}) and (\ref{anom2})
the contribution of $n_2$ cancels.
Similarly, if we sum up all the contribution
from corners of the face for the gauge group $SU(N_1)$,
all contributions cancel,
and we conclude that number assignments satisfying the
D5-brane charge conservation law always give anomaly-free
gauge theories.

In the case of orientifolds, we must impose the $\bZ_2$ invariance
to the numbers assigned to faces.
If a number assignment to faces is $\bZ_2$ invariant, the other
charges assigned to cycles and intersections are also
automatically $\bZ_2$ invariant.
The charge transfer at a fixed point between two cycles intersecting at the fixed point
is always even integer.
Let $M$ be the charge transfer.

If the T-parity of the O5-plane is positive,
the major O5-plane is positive and the minor O5-plane is negative.
Because the flow $M$ is the difference of RR-charges of
the minor part and the major part,
it is given by
\begin{equation}
M=(N_{\rm minor}-2)-(N_{\rm major}+2)
 =N_{\rm minor}-N_{\rm major}-4,
\end{equation}
where $N_{\rm minor}$ and $N_{\rm major}$ are the numbers of branes
coinciding the minor and major O5-planes, respectively.
The chiral multiplets arising at the O5-plane and their contribution $d$ to
the $SU(N)$ anomaly are
\begin{equation}
\symm+N_{\rm major}\fund+N_{\rm minor}\fundbar,\quad
d=(N+4)+N_{\rm major}-N_{\rm minor}
 =N-M.
\label{anompos}
\end{equation}

If the T-parity of the O5-plane is negative,
the major O5-plane is negative and the minor O5-plane is positive.
The charge transfer is given by
\begin{equation}
M=(N_{\rm minor}+2)-(N_{\rm major}-2)
 =N_{\rm minor}-N_{\rm major}+4.
\end{equation}
The chiral multiplets and their contribution to the anomaly are
\begin{equation}
\asymm+N_{\rm major}\fund+N_{\rm minor}\fundbar,\quad
d=(N-4)+N_{\rm major}-N_{\rm minor}
 =N-M.
\label{anomneg}
\end{equation}
Because both (\ref{anompos}) and (\ref{anomneg})
are the same as the anomaly in the parent theory without orientifolding,
the anomaly in the orbifold theory cancels if the parent theory
with the same number assignment
is anomaly free.
Therefore, as the un-orbifolded case,
a charge conserving number assignment
always gives an anomaly-free gauge theory.

%%%%%%%%%%%%%%%%%%%%%%%%%%%%%%%%%%%%%%%%%%%%%%%%%%%%%%%%%%%%%%%%%
\subsection{Examples}\label{example.sec}
Before ending this section, we give a few examples of
fivebrane systems.

\begin{figure}[htb]
\centerline{\scalebox{0.5}{\includegraphics{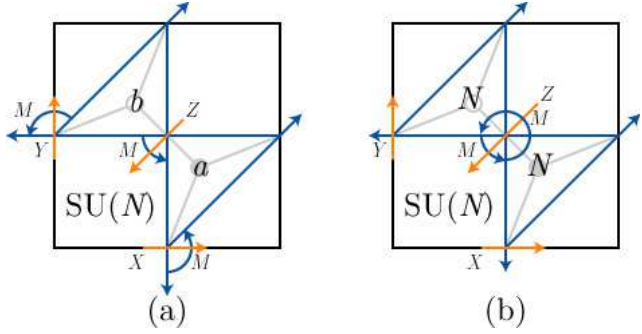}}}
\caption{Two examples of fivebrane diagrams are shown.}
\label{ex2.eps}
\end{figure}

Let us first consider the simplest example,
$\bC^3$ tiling without O5-planes.
((a) in Figure \ref{ex2.eps})
The fivebrane diagram includes three faces.
The hexagonal face with number $N$ being assigned corresponds to gauge group $SU(N)$.
We label the three intersections by $X$, $Y$, and $Z$,
and we use the same letters for the corresponding
bi-fundamental fields.
We assign $a$ and $b$ to the two triangular faces corresponding to the
black and white vertices in the bipartite graph.
The charge conservation uniquely determines
the flow along cycles and the transfer at intersections.
The transfer at three intersections are all equal and given by
\begin{equation}
M=2N-a-b.
\label{threem}
\end{equation}
For simplicity let us assume that $M$ is positive and we introduce only
minor flavor branes.
Then we have the field content in Table \ref{c3fc.tbl}.
\begin{table}[htb]
\caption{The matter content of the gauge theory realized on the
fivebrane system (a) in Figure \ref{ex2.eps}.}
\label{c3fc.tbl}
\begin{center}
\begin{tabular}{ccccc}
\hline
\hline
& $SU(N)$ & $U(M)_X$ & $U(M)_Y$ & $U(M)_Z$ \\
\hline
$X$, $Y$, $Z$ & adj & 1 & 1 & 1 \\
$q_X$ & $\fund$ & $\fundbar$ & 1 & 1 \\
$\wt q_X$ & $\fundbar$ & $\fund$ & 1 & 1 \\
$q_Y$ & $\fund$ & 1 & $\fundbar$ & 1 \\
$\wt q_Y$ & $\fundbar$ & 1 & $\fund$ & 1 \\
$q_Z$ & $\fund$ & 1 & 1 & $\fundbar$ \\
$\wt q_Z$ & $\fundbar$ & 1 & 1 & $\fund$ \\
\hline
\end{tabular}
\end{center}
\end{table}
The superpotential of this theory is
\begin{equation}
W_1=\tr(XYZ-ZYX)
+\tr(\wt q_X X q_X)
+\tr(\wt q_Y Y q_Y)
+\tr(\wt q_Z Z q_Z).
\label{w1}
\end{equation}

By taking T-duality transformation,
this system becomes D7-D3 system
in $\bC^3$.
We can identify the diagonal components of $X$, $Y$, and $Z$
as the position of D3-branes.
Let us focus on one of the D3-branes and treat
the bi-fundamental fields as complex coordinates in $\bC^3$.
The massless loci for quarks are given by $X=0$, $Y=0$, and $Z=0$.
On the web-diagram these are represented as three facets.
The numbers of D7-branes wrapped on the corresponding three divisors
are all equal, and given by (\ref{threem}).
This does not mean that a system with different numbers of D7-branes
wrapped on these divisors is inconsistent.
For example, it is obviously possible to wrap D7-branes
on only one of these divisors.
It is just a stack of parallel D7-branes in the flat spacetime.
The equality of the numbers of D7-branes on the three divisors
is required not for the consistency of the system but for the
possibility to transform the system to a fivebrane system.
This fact implies that if we want to study
all the possible D7-brane configurations in Calabi-Yau cones
by means of brane tilings,
we need to generalize the brane system by including
other kinds of branes.

The second example ((b) in Figure \ref{ex2.eps}) is $\bC^3$
with different flavor branes.
We assign all the faces the same number $N$.
Then the transfer among cycles at each intersection vanishes.
We can, however, introduce the same number of major and flavor branes
at each intersection.
Let us consider the case with $M$ major and $M$ minor flavor branes
at the intersection $Z$.
The field content is shown in Table \ref{c3fc-2.tbl}.
\begin{table}[htb]
\caption{The matter content of the gauge theory
realized on the fivebrane system
in (b) of Figure \ref{ex2.eps}.
$U(M)$ and $U(M)'$ are flavor symmetries realized
as gauge symmetry on major and minor flavor branes, respectively.}
\label{c3fc-2.tbl}
\begin{center}
\begin{tabular}{cccc}
\hline
\hline
& $SU(N)$ & $U(M)$ & $U(M)'$ \\
\hline
$X$, $Y$, $Z$ & adj & 1 & 1 \\
$Q_Z$ & $\fund$ & $\fundbar$ & 1 \\
$\wt Q_Z$ & $\fundbar$ & $\fund$ & 1 \\
$q_Z$ & $\fund$ & 1 & $\fundbar$ \\
$\wt q_Z$ & $\fundbar$ & 1 & $\fund$ \\
\hline
\end{tabular}
\end{center}
\end{table}
The superpotential is given by
\begin{equation}
W_2=\tr(XYZ-ZYX)
+\tr(\wt Q_ZXYQ_Z)
+\tr(\wt q_Z Z q_Z).
\label{w2}
\end{equation}
On the web-diagram, all the three facets
are wrapped by $M$ flavor branes.
The major branes are wrapped on two facets and the minor branes
are wrapped on the other.
This means that by taking T-duality we obtain D7-branes
wrapped on the same divisor as the case of the first example.
The difference between two brane configurations
is Wilson lines on the D7-branes.
In the fivebrane system the position of flavor branes on the $\bT^2$
is different.
This difference is transformed by the T-duality to the difference of
the Wilson lines on the D7-brane worldvolumes.
The effect of the Wilson lines may be interpreted as
a mass deformation of the theory.
We can indeed deform the first example into the second one
by adding the following quark mass term to the superpotential
(\ref{w1}):
\begin{equation}
W_{\rm add}=-\tr(\wt q_Yq_X).
\end{equation}
If we eliminate $q_X$ and $\wt q_Y$ from $W_1+W_{\rm add}$
by using the F-term conditions and identify
$\wt q_X$ and $q_Y$ with $\wt Q_Z$ and $Q_Z$
we obtain the superpotential (\ref{w2}) for the second example.

\begin{figure}[htb]
\centerline{\scalebox{0.5}{\includegraphics{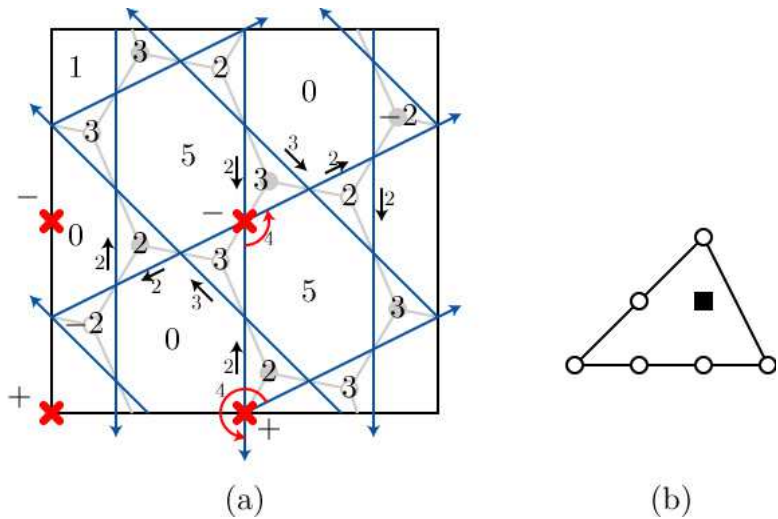}}}
\caption{A fivebrane realization of the supersymmetric Georgi-Glashow model,
which is a famous example of models of dynamical supersymmetry breaking.
(a) is the fivebrane diagram and (b) is the corresponding toric diagram.}
\label{ex1.eps}
\end{figure}

Finally, let us consider an example with O5-planes.
As is pointed out in \cite{Franco:2007ii},
it is easy to make models leading to dynamical supersymmetry breaking
by orientifolded brane tilings.
As examples
they give two brane realizations of the supersymmetric Georgi-Glashow model\cite{Affleck:1983mk}.
Figure \ref{ex1.eps} shows fivebrane diagrams
of one of the models
based on $\bC^3/\bZ_6$ geometry.
This fivebrane diagram includes six hexagonal faces
with integers $0$, $0$, $0$, $1$, $5$, and $5$ assigned to each.
These faces give the gauge group $Sp(0)\times SU(0)\times SO(1)\times SU(5)\sim SU(5)$.
We also have $\ol{\bf 10}$ from the fixed point at the center
and $\bf5$ from the contacting point of the $SU(5)$ face and the $SO(1)$ face.
The D5-brane charge is conserved without introducing any flavor branes,
and we have no more extra fields.

%%%%%%%%%%%%%%%%%%%%%%%%%%%%%%%%%%%%%%%%%%%%%%%%%%%%%%%%%%%%%%%
\section{Conclusions and discussions}\label{discussions.sec}
In this paper we investigated orientifold
of brane tilings from the perspective of
fivebrane systems.
Among several possibilities, we only discussed
orientifold with O5-planes
which are represented as points in bipartite graphs.
We showed that
the cancellation of Witten's $\bZ_2$ anomaly and
gauge anomaly are guaranteed by the conservation of
D5-brane charge.
The fact that the O5-plane RR-charge flips at the intersection
with NS5-brane plays the key role to explain the gauge anomaly cancellation.
The charge conservation requires the emergence of appropriate
flavor branes,
which do not necessarily coincide with the O5-planes, to
compensate for the change of the O5-plane charges.
These flavor branes provide quark fields
which cancel the gauge anomaly generated by symmetric
or antisymmetric representation fields.

In section \ref{cy.sec} we investigated the
relation between fivebrane systems to Calabi-Yau cones.
We gave the formula for $\bZ_2$ parity of mesonic operators
in terms of RR-charges
of O5-planes, and establish the relation
between RR-charge and T-parity.
We also gave quark mass terms
which reproduce the world volumes of
flavor branes in toric Calabi-Yau
cones as the massless loci
of quark fields.

All these are quite remarkable, but still there are many open questions.

As we mentioned above, we only investigated the case of orientifold with
O5-planes.
There are, however, other possibilities with O7-planes.
The matter contents for such case is given in \cite{Franco:2007ii},
and similarly to the O5-brane case,
we need fundamental matter fields to cancel gauge anomaly.
The emergence of flavor branes should be guaranteed by some
consistency in the brane system and it is important to
clarify how the quark fields arise in these models.

In \S\ref{cycone.sec}, we restrict our attention
only to RR-charge assignments with positive total charge.
(Total charge is the product of RR-charges of four O5-planes.)
From the fivebrane perspective,
it seems also possible to consider RR-charge assignments with
negative total charge.
These two possibilities can be distinguished by
two $\bZ_2$ invariant value of the integral
\begin{equation}
b=\int_{T^2}B_2.
\end{equation}
The invariance under the orientifold flip $b\rightarrow -b$
restrict this value to be $0$ or $\pi$ mod $2\pi$.
As is clarified in \cite{Imamura:2007dc}, this parameter
is related to the $\beta$-deformation
in the gauge theory.
It may be interesting to investigate the relation
between the brane configurations with negative total RR-charge
and $\beta$-deformation in gauge theories.

In general, the gauge theories realized by
the fivebrane systems are not conformal.
Such gauge theories without conformal symmetry are known to
enjoy phenomenon so-called duality cascade\cite{Klebanov:2000hb}.
It is known\cite{Elitzur:1997fh}
that type IIA brane construction of
${\cal N}=1$ gauge theories provides
simple way to realize Seiberg duality\cite{Seiberg:1994pq}
in a geometric way.
It would be interesting to study the phenomenon by using
the fivebrane systems investigated in this paper.

Construction of models of dynamical SUSY breaking
from orientifolded brane tilings is an extremely interesting issue,
considering its phenomenological interest.
Our analysis clarified the structure of the fivebrane system realizing
the model proposed in \cite{Franco:2007ii}.
We need to consider non-trivial flow of flavor D5-brane charges
along NS5-branes.
Alternatively, we can use metastable SUSY breaking\cite{Intriligator:2006dd}.
In un-orientifolded case it is shown that we have metastable vacuum
by the inclusion of flavor branes\cite{Franco:2006es,GarciaEtxebarria:2007vh},
and we expect existence of metastable vacua in orientifolded case as well.

Finally, it would be interesting to consider application
to mirror symmetry for Calabi-Yau orientifolds,
since in the un-orientifolded case, brane tilings
are quite useful in proving homological mirror
symmetry\cite{UY1,UY2,UY3}.
We expect several differences,
such as that $A_{\infty}$-structures are
replaced by $L_{\infty}$-structures
in orientifold case\cite{Diaconescu:2006id},
but the basic line of argument should be similar.

%%%%%%%%%%%%%%%%%%%%%%%%%%%%%%%%%%%%%%%%%%%%%%%%%%%%%%%%%%%%%%%%%

We hope to return to these topics in the future.

\subsection*{Acknowledgements}
We would like to thank F. Koyama for valuable discussions.
Y.~I. would also like to thank Osaka City University Research Group for Mathematical Physics
for the oppertunity of giving a lecture about brane tilings.
Y.~I. is partially supported by
Grant-in-Aid for Young Scientists (B) (\#19740122) from the Japan
Ministry of Education, Culture, Sports,
Science and Technology.

\appendix

%%%%%%%%%%%%%%%%%%%%%%%%%%%%%%%%%%%%%%%%%%%%%%%%%%%
\section{Proofs of theorems}
\subsection{Some identifies}\label{eqns.sec}
In this sections we prove three formulae,
originally labeled (\ref{ocu5}), (\ref{ocu12}), and (\ref{ocu13}).
The first formula (\ref{ocu5}) is
\begin{equation}
\sigma(\rho(Z_{\alpha+1,\alpha}))=h(Z_{\alpha+1,\alpha})
\mod2 .
\label{formula1}
\end{equation}
The second formula (\ref{ocu12})
says if $C$ is a $\bZ_2$ symmetric path,
\begin{equation}
(-1)^{\langle m_\alpha,C[{\cal O}]\rangle}
=\int_{C[{\cal O}]}\rho(m_\alpha).
\label{formula2}
\end{equation}
The final formula (\ref{ocu13}) is
\begin{equation}
(-1)^{\langle m[s_\alpha],C[{\cal O}]\rangle}
=\int_{C[{\cal O}]}Q_\alpha.
\label{formula3}
\end{equation}
We have renumbered these formulae for convenience.

We first prove (\ref{formula2}). By definition,
$\langle m_\alpha,C[{\cal O}]\rangle$ is given
by the intersection number of $C[{\cal O}]$
with the perfect matching $m_{\alpha}$.
In general, we have many possible intersection points,
but if you consider $\mod2$ intersection number,
only intersection points on orientifold planes contribute
because other intersection points always come in pairs. We thus have 
\begin{equation}
(-1)^{\langle m_\alpha,C[{\cal O}]\rangle}
=(-1)^{ (\textrm{number of intersection points of $C[{\cal O}]$ with
$m_\alpha$ at fixed points})}.
\end{equation}
The RHS is nothing but the definition of $\int_{C[{\cal O}]} \rho(m_{\alpha})$, which proves (\ref{formula2}). 

Now the proof of other formulae are easy.
(\ref{formula1}) is shown from (\ref{formula2}) as follows:
\begin{equation}
\begin{split}
(-1)^{\sigma(\rho(Z_{\alpha+1,\alpha}))_1}
&= \int_{\balpha} \rho (Z_{\alpha+1,\alpha})
\quad (\since (\ref{-1sQ}))\\
&= \int_{\balpha} \rho (m_{\alpha+1}) \int_{\balpha} \rho(m_{\alpha})
\quad (\since (\ref{Zmm}))\\
&= (-1)^{\langle m_{\alpha+1}, \balpha \rangle}
   (-1)^{\langle m_{\alpha}, \balpha \rangle}\quad
(\since (\ref{formula2}) \textrm{ with  $C[{\cal O}]=\balpha$-cycle}) \\
&= (-1)^{\langle Z_{\alpha+1,\alpha}, \balpha \rangle}
\quad (\since (\ref{ocu7}))\\
&=(-1)^{h(Z_{\alpha+1,\alpha})_1}
\quad (\since (\ref{heightdef})) .
\end{split}
\end{equation}
(The subscripts $1$ mean the first component of the vectors.)

Similarly, when $C[{\cal O}]$ is the $\balpha$-cycle, we have
\begin{equation}
\begin{split}
(-1)^{\langle m[s_\alpha],C[{\cal O}]\rangle} 
& = (-1)^{\langle m[s_\alpha],\balpha \rangle} \\
&=  (-1)^{h(m[s_\alpha])_1}
\quad (\since (\ref{heightdef}))\\
&=  (-1)^{{s_{\alpha}}_1}
\quad(\since (\ref{mviab}) \textrm{ and } (\ref{ocu3})) \\
&= \int_{\balpha} Q_{\alpha}
\quad (\since (\ref{-1sQ})).
\end{split}
\end{equation}
The case of more general $C[{\cal O}]$ is similar,
and this proves (\ref{formula3}).

%%%%%%%%%%%%%%%%%%%%%%%%%%%%%%%%%%%%%%%%%%%%%%%%%%%%%%%%%%%%%%%%%%%%%%
\subsection{Divisors and GLSM fields}\label{GLSM.sec}
We here prove theorem \ref{divisor}.
Namely, we show that in the subspace ${\cal M}'$
defined in \S\ref{qmt.sec}
the divisor $F_\alpha$ is given by $\rho_\alpha=0$,
where $\rho_\alpha$ is the GLSM field corresponding to the
unique perfect matching $m_\alpha$ for the corner $\alpha$ of the toric diagram.
For any perfect matching $m_{\alpha'}$, which does not necessarily correspond
to a corner of the toric diagram, there is a corresponding
GLSM field $\rho_{\alpha'}$,
and in terms of these GLSM fields
the solution of the F-term condition
is given by\cite{Franco:2006gc}
\begin{equation}
\Phi_I=\prod_{\alpha'\ni I}\rho_{\alpha'},
\label{glsm}
\end{equation}
where $\alpha'\ni I$ means all perfect matchings
including the edge $I$.

The system of $\rho_{\alpha'}$ does not have superpotential
and possesses $U(1)^n$ symmetry,
where $n$ is the number of the GLSM fields.
Among these $U(1)$ symmetries, $n-3$ are gauged.
The gauged subgroup $U(1)^{n-3}\subset U(1)^n$ is specified by
the charge matrix $g_k^{\alpha'}$, $k=1,\ldots,n-3$.
The gauge transformation of $\rho_{\alpha'}$
with parameter $\theta_k$ are given by
\begin{equation}
\rho'_{\alpha'}=e^{i\sum_k\theta_kg_k^{\alpha'}}\rho_{\alpha'}.
\end{equation}
By removing unphysical degrees of freedom
associated with (complexified version of) this gauge symmetry,
we obtain the three dimensional moduli space.
We denote the natural map from $\rho$ space ${\cal M}_{\rm GLSM}$ to
the three dimensional moduli space ${\cal M}$ by $\psi$:
\begin{equation}
\psi:{\cal M}_{\rm GLSM} \rightarrow {\cal M}.
\label{ocu16}
\end{equation}

These gauge transformation generate shifts in the space of
angular variables $\varphi_{\alpha'}=\arg\rho_{\alpha'}$,
and the toric fiber $\bT^3$ of the moduli space ${\cal M}$
can be regarded as classes defined by the
following identification relation:
\begin{equation}
\varphi_{\alpha'}\sim \varphi_{\alpha'}+\theta_kg_k^{\alpha'}.
\end{equation}
We also use $\psi$ for the natural homomorphism associated with
this relation.
Then, the points in the toric diagram are related with
the unit vectors in the $\varphi_{\alpha'}$ space $e_{\alpha'}$ by
\begin{equation}
v_{\alpha'}=\psi(e_{\alpha'}).
\end{equation}
By this relation, we can relate the symmetry in the
GLSM and that of the moduli space ${\cal M}$.
The symmetry $U(1)[e_{\alpha'}]$ acting on ${\cal M}_{\rm GLSM}$ induces the
isometry $U(1)[v_{\alpha'}]$ of the moduli space.

We will now prove the theorem \ref{divisor}:
\begin{itemize}
\item
In the subspace ${\cal M}'\subset{\cal M}$ a divisor $F_\alpha$ is given by $\rho_\alpha=0$.
\end{itemize}
The restriction to ${\cal M}'$ means that
we neglect the subspace corresponding to the legs and the center of
the web-diagram as we mention in \S\ref{d7.sec}.

It is obvious that if $\rho_\alpha=0$ the corresponding points
in the moduli space is in the divisor $F_\alpha$ because
$\rho_\alpha=0$ is a fixed point of the symmetry $U(1)[e_\alpha]$,
which induce $U(1)[v_\alpha]$ in the moduli space.
What is slightly non-trivial is the converse of this statement.
Namely, we need to show that
\begin{itemize}
\item
If $\psi(\rho)$ is a point inside a divisor $F_\alpha$, then $\rho_\alpha=0$.
\end{itemize}
We assume that $\psi(\rho)\in {\cal M'}$, and
$\psi(\rho)$ is not shared by more than one facets.

Let us choose one corner $\alpha_0$ in the toric diagram, and
assume that $\psi(\rho)$ is a point inside a divisor $F_{\alpha_0}$.
This means that
$\rho$ is invariant under $U(1)[e_{\alpha_0}]$ up to gauge transformation.
\begin{equation}
e^{i\theta\delta_{\alpha_0}^{\beta'}}\rho_{\beta'}
=e^{i\theta\sum_kc_kg_k^{\beta'}}\rho_{\beta'},\quad
(\forall \theta,{\beta'},\ 
\exists c_k) .
\label{ocu17}
\end{equation}
One trivial solution is
\begin{equation}
\rho_{\alpha_0}=c_i=0.
\label{ocu175}
\end{equation}
We want to show that this is only solution to (\ref{ocu17}).

Let us assume $\rho_{\alpha_0}\neq0$.
The relation (\ref{ocu17}) with $\beta'=\alpha_0$ requires
\begin{equation}
\sum_kc_kg_k^{\alpha_0}=1.
\label{ocu18}
\end{equation}
Because $\sum_{\alpha'} g_k^{\alpha'}=0$ for all $k$
due to the GLSM $U(1)$ anomaly cancellation condition,
there is at least one $\beta'\neq\alpha_0$ with which
\begin{equation}
\sum_kc_kg_k^{\beta'}\neq0.
\label{ocu19}
\end{equation}
If there is only one such a $\beta'$, say $\alpha'_1$,
it satisfy
\begin{equation}
\sum_kc_kg_k^{\alpha'_1}=-1.
\label{ocu20}
\end{equation}
From (\ref{ocu18}) and (\ref{ocu20}), we have
\begin{equation}
\delta_{\alpha_0}^{\beta'}-\delta_{\alpha'_1}^{\beta'}=\sum_kc_kg_k^{\beta'},
\quad
(\forall\beta').
\label{ocu21}
\end{equation}
This relation, however,
means that $\psi(e_{\alpha_0})=\psi(e_{\alpha'_1})$, and
the two GLSM fields $\rho_{\alpha_0}$ and $\rho_{\alpha'_1}$,
and thus two perfect matchings $m_{\alpha_0}$ and $m_{\alpha'_1}$,
correspond to the same point at a corner in the toric diagram.
This contradicts our assumption.

If there are more than one $\beta'$,
say $\alpha'_1$, $\alpha'_2,\ldots$,
satisfying (\ref{ocu19}),
then from (\ref{ocu17}), we have
\begin{equation}
\rho_{\alpha'_1}=\rho_{\alpha'_2}=\cdots=0.
\end{equation}
This means that $\psi(\rho)$ shared by more than one divisors
and again contradicts our assumption that $\psi(\rho)$
is a point inside the divisor $F_{\alpha_0}$.
Therefore, (\ref{ocu175}) is the only
solution to the
relation (\ref{ocu17}).

\subsection{Theorem \ref{minor}} \label{thm.sec}
Let us prove the theorem \ref{minor}.

The statement that the set of facets form one continuous region can be shown by using the
fact that edge $I$ is always included in two and only two
non-parallel zig-zag paths.
This means that $m_{\alpha+1}[I]-m_\alpha[I]$ becomes non-zero
for two $\alpha$, and they give the boundaries which divide
the plane of web-diagram into two parts.

Let $m_{\alpha+1}-m_\alpha$ and $m_{\beta+1}-m_\beta$
be the two zig-zag paths
which give the boundaries.
There are two possibilities
\begin{itemize}
\item case (i) :
$m_{\alpha+1},m_\beta\ni I$,
$m_\alpha,m_{\beta+1}\notni I$.
\item case (ii) :
$m_{\alpha+1},m_\beta\notni I$,
$m_\alpha,m_{\beta+1}\ni I$.
\end{itemize}

\begin{figure}[htb]
\centerline{\scalebox{0.5}{\includegraphics{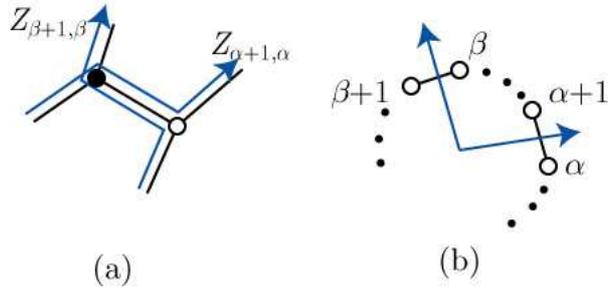}}}
\caption{(a) shows two zig-zag paths passing through an edge.
(b) shows the corresponding external legs in the web-diagram.}
\label{ocu.eps}
\end{figure}

In the case (i),
because $m_{\alpha+1}$ includes $I$,
$Z_{\alpha+1,\alpha}=m_{\alpha+1}-m_\alpha$
passes the edge $I$ from black to white,
while $Z_{\beta+1,\beta}$ passes in the opposite direction,
from white to black.
By the definition of zig-zag paths,
we see that $Z_{\beta+1,\beta}$ crosses $Z_{\alpha+1,\alpha}$
upward when $Z_{\alpha+1,\alpha}$ goes left to right.
(Figure \ref{ocu.eps} (a))
This implies that the facets $F_{\beta}$ and $F_{\alpha+1}$ are
on the side of the minor angle
made by two legs (Figure \ref{ocu.eps} (b)),
and the all facets whose perfect matchings include the edge $I$ are also
on the same side.
The same is shown for the case (ii) and the proposition has been proved.

%%%%%%%%%%%%%%%%%%%%%%%%%%%%%%%%%%%%%%%%%%%%%%%%%%%%%%%%%%%%%
\end{document}